\documentclass[11pt]{article}
\usepackage{amsmath,amssymb,color,epsfig,cite}
\usepackage{graphicx}
\usepackage{subfigure}
\usepackage{setspace}
\usepackage{epstopdf}

\usepackage{amsfonts}

\newcommand{\hoch}[1]{$\, ^{#1}$}

\makeatletter
\@addtoreset{equation}{section}
\makeatother

\newcommand{\be}{\begin{equation}}
\newcommand{\ee}{\end{equation}}
\newcommand{\bea}{\setlength\arraycolsep{2pt} \begin{eqnarray}}
\newcommand{\eea}{\end{eqnarray}}

\def\ft#1#2{{\textstyle{\frac{\scriptstyle #1}{\scriptstyle #2} } }}
\def\fft#1#2{{\frac{#1}{#2}}}

\def\0{{\sst{(0)}}}
\def\1{{\sst{(1)}}}
\def\2{{\sst{(2)}}}
\def\3{{\sst{(3)}}}
\def\4{{\sst{(4)}}}
\def\5{{\sst{(5)}}}
\def\6{{\sst{(6)}}}
\def\7{{\sst{(7)}}}
\def\8{{\sst{(8)}}}
\def\9{{\sst{(9)}}}

\def\sst#1{{\scriptscriptstyle #1}}

\begin{document}



\begin{center}
{\large {\bf More on Heavy-Light Bootstrap up to Double-Stress-Tensor
}}

\vspace{10pt}
Yue-Zhou Li\hoch{1\Delta, 2 J}, Hao-Yu Zhang\hoch{3 \gamma}

\vspace{15pt}

{\it \hoch{1}Center for Joint Quantum Studies and Department of Physics,\\
School of Science, Tianjin University, Tianjin 300350, China \\
{\it \hoch{2} Department of Physics, McGill University, 3600 Rue University, Montr\'{e}al, QC Canada\\
{\it \hoch{3} George P. \& Cynthia Woods Mitchell  Institute
for Fundamental Physics and Astronomy,\\
Texas A\&M University, College Station, TX 77843, USA}}}

\vspace{30pt}

\underline{ABSTRACT}
\end{center}
We investigate the heavy-light four-point function up to double-stress-tensor, supplementing 1910.06357. By using the OPE coefficients of lowest-twist double-stress-tensor in the literature, we find the Regge behavior for lowest-twist double-stress-tensor in general even dimension within the large impact parameter regime. In the next, we perform the Lorentzian inversion formula to obtain both the OPE coefficients and anomalous dimensions of double-twist operators $[\mathcal{O}_H\mathcal{O}_L]_{n,J}$  with finite spin $J$ in $d=4$. We also extract the anomalous dimensions of double-twist operators with finite spin in general dimension, which allows us to address the cases that $\Delta_L$ is specified to the poles in lowest-twist double-stress-tensors where certain double-trace operators $[\mathcal{O}_L\mathcal{O}_L]_{n,J}$ mix with lowest-twist double-stress-tensors. In particular, we verify and discuss the Residue relation that determines the product of the mixed anomalous dimension and the mixed OPE. We also present the double-trace and mixed OPE coefficients associated with $\Delta_L$ poles in $d=6,8$. In the end, we turn to discuss CFT$_2$, we verify the uniqueness of double-stress-tensor that is consistent with Virasoso symmetry.

\vfill {\footnotesize \hoch{\Delta}liyuezhou@tju.edu.cn\ \ \ \ \hoch{J}liyuezhou@physics.mcgill.ca \ \ \ \ \hoch{\gamma}haoyuzhang001@gmail.com}

\pagebreak

\tableofcontents
\addtocontents{toc}{\protect\setcounter{tocdepth}{2}}


\newpage
\section{Introduction}
\label{intro}
A conformal field theory (CFT) is characterized by the conformal dimensions, spin and operator product expansion (OPE) coefficients of all existed local primary operators. Remarkably, by taking advantage of the conformal symmetry combined with the general consistency conditions such as the unitarity and the crossing equation, conformal bootstrap enables us to explore the conformal dimensions and OPE coefficients in a powerful and efficient way. For example, the numerical bootstrap sets the crossing equation as the semi-definite-programming, providing the powerful ability to narrow down the possible value of conformal dimensions and OPE coefficients up to high precision, e.g. see \cite{Poland:2018epd} for review.

In parallel to the numerical bootstrap, the analytic bootstrap, in particular, the lightcone bootstrap makes use of the singularity near the lightcone limit (in the Lorentzian signature) of the crossing equation, from which the large spin operators \cite{Alday:2007mf} arises naturally and were analyzed extensively, e.g. \cite{Fitzpatrick:2012yx,Komargodski:2012ek,Alday:2013cwa,Kaviraj:2015xsa}. Generally speaking, the CFT data of the large spin operators can be explored by asymptotically expanding them in terms of the inverse powers of spin $1/J$ and subsequently solving the singular part of crossing equation order by order algebraically \cite{Kaviraj:2015cxa,Alday:2015ewa}. The large spin perturbation theory was then developed to extract the large spin data up to all orders of $1/J$ \cite{Alday:2016njk}. The CFT data bootstrapped from this analytic procedure are actually valid to all spin except for few low spins $J=0,1$ \cite{Alday:2015ota,Simmons-Duffin:2016wlq}, which can be explained by the analyticity in spin explicitly from Lorentzian inversion formula \cite{Caron-Huot:2017vep,Simmons-Duffin:2017nub,Kravchuk:2018htv}. By employing these highly-developed techniques of lightcone bootstrap, considerable progress was made in numerous subjects, for example, Wilson-Fisher and $O(N)$ models \cite{Alday:2017zzv,Alday:2019clp,Henriksson:2018myn} and AdS supergravity \cite{Alday:2017xua,Alday:2017vkk,Caron-Huot:2018kta}.

Recently, the four-point function with heavy states in large $N\sim \sqrt{C_T}$ CFT where the heavy conformal dimension is comparable to $C_T$ charge, known as heavy-light four-point function, draws a lot of attentions. In $d=2$, the heavy-light four-point function is universally constructed by the Virasoro identity block \cite{Fitzpatrick:2014vua} and enjoys plentiful applications in the context of AdS/CFT \cite{Maldacena:1997re,Gubser:1998bc,Witten:1998qj}, e.g. \cite{Fitzpatrick:2014vua,Fitzpatrick:2015zha,Anous:2016kss,Hartman:2013mia,Asplund:2014coa}. It is natural to expect that higher dimensional CFT may have similarity to CFT$_2$ within a certain kinematic limit, for example, the Virasoro-like structure can be observed near the lightcone limit \cite{Huang:2019fog,Huang:2020ycs}. The heavy-light four-point function would be an appropriate window to probe the similarity. It is indeed found from holographic set-ups that the universal piece of heavy-light four-point function is the lowest-twist multi-stress-tensors \cite{Fitzpatrick:2019zqz,Fitzpatrick:2019efk}, similar to the universality in CFT$_2$ imposed by Virasoro symmetry. Thus it would be interesting and important to utilize the developed lightcone bootstrap to study heavy-light four-point function from which the underlying CFT data can be extracted \cite{Kulaxizi:2019tkd,Karlsson:2019dbd,Li:2019zba,Karlsson:2020ghx} and the evidence of the universality can be provided and understood \cite{Li:2019zba}. This paper is a supplement to the previous paper \cite{Li:2019zba} that bootstraps the heavy-light four-point function by using the Lorentzian inversion formula. We discuss the large impact parameter regime of the Regge limit \cite{Costa:2012cb} for lowest-twist double-stress-tensor in the heavy-light four-point function, and more importantly, we extend the universality of the double-twist operators $[\mathcal{O}_H\mathcal{O}_L]_{n,J}$ at the leading order $\mathcal{O}(\mu)$ from large spin to finite spin and then tackle the $\Delta_L$ poles referred in \cite{Fitzpatrick:2019zqz,Fitzpatrick:2019efk,Li:2019tpf,Li:2019zba}. We also discuss the case in $d=2$, showing the consistency with Virasoro identity block \cite{Fitzpatrick:2014vua}.

The paper is organized as follows: in section \ref{review}, we review the heavy-light four-point function and the recent progress on this subject and then we argue that there is no mixing problem in the heavy-light bootstrap; in section \ref{regge}, we analyze the Regge behavior of lowest-twist double-stress-tensor where a general even dimensional expression in the large impact parameter regime is provided; in section \ref{finitej}, we employ the Lorentzian inversion formula to extract the OPE coefficients and anomalous dimension of double-twist operators $[\mathcal{O}_H\mathcal{O}_L]_{n,J}$ with finite spin in $d=4$, and moreover, we also manage to find the anomalous dimension in any dimension $d$ by using $\bar{z}\rightarrow 1$ expansion of the logarithmic part of the stress-tensor conformal block we found; in section \ref{deltapole}, we consider the case that $\Delta_L$ approaches the poles in the lowest-twist double-stress-tensor OPE and verify the Residue relation proposed in \cite{Li:2019tpf}, the mixed OPE coefficients in $d=4$ is obtained and the universality of the Residue relation is discussed; in section \ref{CFTtwo}, we consider $d=2$ and obtain the double-twist data with finite spin. By using these data, we then find the uniqueness of double-stress-tensor, which is consistent with Virasoro symmetry; the paper is summarised in section \ref{conclu}; the Lorentzian inversion formula is reviewed in Appendix \ref{inversion} and the series expansion of conformal blocks is reviewed in Appendix \ref{seriesexp}; in Appendix \ref{geowitten}, we briefly explain how we derive $\bar{z}\rightarrow 1$ expansion of the logarithmic part of the stress-tensor conformal block from geodesic Witten diagram; in Appendix \ref{holography}, we show how to derive the answer of anomalous dimension of $[\mathcal{O}_H\mathcal{O}_L]_{n,J}$ with finite spin from holography by using the holographic Hamiltonian perturbation theory; in Appendix \ref{mixOPE}, we consider all examples of $\Delta_L$ poles in $d=6,8$ and work out the OPE coefficients of double-trace operators $[\mathcal{O}_L\mathcal{O}_L]_{n,J}$ up to the mixed twist.
\section{Review of the heavy-light bootstrap}
\label{review}
In this section, we review the set-up and the recent progress of bootstrapping heavy-light four-point function, moreover, we would like to comment on the mixing problem of heavy-light bootstrap mentioned in \cite{Li:2019zba} and argue that we do not need to worry about the mixing problem.

\subsection{The heavy-light four-point function}
We consider a generic large $N\sim \sqrt{C_T}$ CFT with a large gap $\Delta_{\rm gap}$ such that it admits holographic gravity dual \cite{Heemskerk:2009pn}. The objects we are interested in bootstrapping is the heavy-light four-point function $\langle \mathcal{O}_H\mathcal{O}_L\mathcal{O}_L\mathcal{O}_H\rangle$ in such a theory, which consist of two heavy operators with conformal dimension $\Delta_H \sim C_T$ and two light operators with conformal dimension $\Delta_L \ll C_T$. The heavy-light four-point function can be studied from both $s$-channel ($\mathcal{O}_H\mathcal{O}_L\rightarrow \mathcal{O}_L\mathcal{O}_H$) and $t$-channel ($\mathcal{O}_H\mathcal{O}_H\rightarrow \mathcal{O}_L\mathcal{O}_L$), and the crossing equation can be established within an appropriate conformal frame
\be
(z\bar{z})^{\fft{\Delta_H+\Delta_L}{2}}\langle \mathcal{O}_H\mathcal{O}_L\mathcal{O}_L\mathcal{O}_H\rangle=\mathcal{G}^s(z,\bar{z})=
\fft{(z\bar{z})^{\fft{\Delta_H+\Delta_L}{2}}}{((1-z)(1-\bar{z}))^{\Delta_L}}\mathcal{G}^t(1-\bar{z},1-z)\,,\label{crossingeq}
\ee
where both $s$ and $t$-channel correlators $\mathcal{G}^{s,t}$ shall be expanded in terms of $\mu\sim \Delta_H/C_T$ around the degenerate point $\mu\rightarrow 0$\footnote{Note this is not contradict to the heavy-limit $\Delta_L\ll\Delta_H\sim C_T$ we consider. We can take the heavy-limit $\Delta_L\ll\Delta_H\sim C_T$ for correlators at first such that the correlators are functions of $\mu$ and in general they are not polynomials, for example, see examples in CFT$_2$ \cite{Fitzpatrick:2014vua}. Then we can expand the resulting correlators in terms of $\mu$ around $\mu\rightarrow 0$. Another perspective is that we can start with large $N$ expansion where the degenerate point is located at $C_T\rightarrow \infty$ and then we collect $\Delta_H$ to reorganize the expansions in terms of $\mu$ \cite{Li:2019zba}.}. To be more precise, we adopt the convention in \cite{Kulaxizi:2018dxo}
\be
\mu=\frac{4 \Gamma(d+2)}{(d-1)^2\Gamma(\frac{d}{2})^2}\frac{\Delta_H}{C_T}\,.
\ee
At the degenerate point, the heavy-light four-point function is the four-point function of generalized free field theory, thus $\mathcal{G}^t=1$ and the degenerate double-twist operators $[\mathcal{O}_H\mathcal{O}_L]_{n,J}=\mathcal{O}_L\partial^{2n}\partial_{\mu_1}\cdots\partial_{\mu_J}\mathcal{O}_L$ with conformal dimensions $\Delta=\Delta_H+\Delta_L+J+2n$ are exchanged in $s$-channel weighted by the free OPE coefficients \cite{Fitzpatrick:2011dm}, where heavy-limit should be imposed
\be
\tilde{c}_{n,J}^{\textrm{free}}=\frac{(J+1)\Gamma(\Delta_L+n-1)\Gamma(\Delta_L+n+J)}{\Gamma(n+1)\Gamma(n+J+2)\Gamma(\Delta_L-1)\Gamma(\Delta_L)}
+\mathcal{O}(\frac{1}{\Delta_H})\,.\label{freeOPEheav}
\ee
In general, the degeneracy of the double-twist operators breaks down via acquiring anomalous dimensions $\Delta_L=\Delta_H+\Delta_L+J+2n+\tilde{\gamma}_{n,J}$ and the corresponding OPE coefficients are corrected to be $\tilde{c}_{n,J}$ where
\bea
\tilde{c}_{n,J}=\tilde{c}_{n,J}^{\textrm{free}}\sum_{k=0} \mu^k \tilde{c}_{n,J}^{(k)},\qquad \tilde{\gamma}_{n,J}=\tilde{c}_{n,J}^{\textrm{free}}\sum_{k=1} \mu^k \tilde{\gamma}_{n,J}^{(k)}\,,
\eea
where it turns out $\tilde{\gamma}^{(k)}$ and $\tilde{c}^{(k)}$ are universal at large spin limit and behave as \cite{Kulaxizi:2018dxo,Karlsson:2019qfi,Kulaxizi:2019tkd,Karlsson:2019dbd,Li:2019zba}
\be
\tilde{c}_{n,J}^{(k)}\, \,  \textrm{and} \, \, \tilde{\gamma}_{n,J}^{(k)} \sim (J')^{-\frac{k(d-2)}{2}}
\ee

By dimensional analysis, the operators exchanged in $t$-channel shall be multi-stress-tensors $T^k$ (the multi-twist operators constructed from stress-tensor) living at the order $\mathcal{O}(\mu^k)$
\be
T \partial\cdots T \partial\cdots T \cdots\,,
\ee
for example, at $\mathcal{O}(\mu)$ it is assumed that there is only single-stress-tensor with OPE coefficient completely determined by Ward identity \cite{Osborn:1993cr,Dolan:2000ut}
\be
c_{T}=\frac{\Delta_L \Gamma(\frac{d}{2}+1)^2}{4 \Gamma(d+2)} \mu\,,\label{TOPE}
\ee
which is universal by depending only on $\Delta_H,\Delta_L,C_T$. In general, as the convention used for $s$-channel, we usually factorize $\mu$ out and denote
\be
c_{T^k(\Delta,J)}=\mu^k c_{T^k}\,,\qquad \Delta-J=kd-J_T+2n\,,\qquad J_T\leq 2k \,,\qquad J\geq J_T\,,
\ee
where $J_T$ is the spin of stress-tensors.

Recently, investigating heavy-light four-point function in both $s$ and $t$-channel becomes increasingly attractive, driven by the motivation of looking for the Virasoro-like structures in higher dimensional CFT. In fact, both the holography and the bootstrap allow us to explore $\tilde{c}^{(k)}_{n,J}$, $\tilde{\gamma}^{(k)}_{n,J}$ and $c_{T^k}$. The holography treats $s$ and $t$-channel separately by using different holographic techniques. In $s$-channel, one can use the bulk phase shift \cite{Cornalba:2006xk,Cornalba:2006xm,Cornalba:2007zb,Kulaxizi:2017ixa} or holographic Hamiltonian perturbation theory \cite{Kaviraj:2015xsa,Fitzpatrick:2014vua} to study the large spin limit of $\tilde{c}^{(k)}_{n,J}, \tilde{\gamma}^{(k)}_{n,J}$ \cite{Kulaxizi:2018dxo,Karlsson:2019qfi}; in $t$-channel, the formalism developed in \cite{Fitzpatrick:2019zqz} deals with heavy operators by assuming they form a black hole and consequently one is allowed to compute $c_{T^k}$ by analyzing two-point functions under black holes \cite{Fitzpatrick:2019zqz,Fitzpatrick:2019efk,Li:2019tpf}. The holographic treatments surprisingly suggest that the large spin double-twist operators $[\mathcal{O}_H\mathcal{O}_L]_{n,J}$ and the lowest-twist multi-stress-tensors (i.e. $n=0$ and $J_T=2k$) are all universal where it turns out the higher derivative gravity terms play no role. In addition, the universality in Regge limit can also be investigated from holography \cite{Fitzpatrick:2019efk,Karlsson:2019txu}.

From pure CFT perspective, the bootstrap handles $s$ and $t$-channel within the same framework: the crossing equation (\ref{crossingeq}) enables us to understand both $s$ and $t$-channel. By establishing the ansatz of multi-stress-tensor conformal blocks and assuming large spin expansion of double-twist operators $[\mathcal{O}_H\mathcal{O}_L]_{n,J}$, the crossing equation (\ref{crossingeq}) can then be solved algebraically order by order in $\mu$ from which the large spin limit of $\tilde{c}^{(k)}_{n,J}, \tilde{\gamma}^{(k)}_{n,J}$ and lowest-twist parts of $c_{T^k}$ can be obtained \cite{Karlsson:2019dbd}. With more general ansatz around the lightcone limit, both universal parts and non-universal parts of $1/J$ corrections in $\tilde{c}^{(k)}_{n,J}, \tilde{\gamma}^{(k)}_{n,J}$ and OPE coefficients of higher twist multi-stress-tensors $c_{T^k}$ can also be investigated \cite{Karlsson:2020ghx}. In fact, it turns out even the ansatz of stress-tensor conformal blocks and the assumption of large spin expansion are not necessary in bootstrapping heavy-light four-point function and understanding the hidden universality (though, the efficiency is largely promoted provided with those ansatz and assumption, see \cite{Karlsson:2020ghx}), the Lorentzian inversion formula can be used back and forth to provide the strong evidence that the double-twist operators at large spin limit and lowest-twist multi-stress-tensors show universality, strikingly bootstrapping the large spin limit of $\tilde{c}^{(k)}_{n,J}, \tilde{\gamma}^{(k)}_{n,J}$ and the lowest-twist OPE coefficients $c_{T^k}$ from nothing more than single-stress-tensor OPE (\ref{TOPE}) \cite{Li:2019zba}. The extracted OPE coefficients and anomalous dimensions achieve the exact agreements with holographic set-ups \cite{Kulaxizi:2018dxo,Karlsson:2019qfi,Fitzpatrick:2019efk}. However, some questions quoted in \cite{Li:2019zba}, e.g. $\Delta_L$ poles, mixing problems, etc, still remain unclear.

To go ahead, in this paper, we mainly follow \cite{Li:2019zba}, using the Lorentzian inversion formula to compute the finite spin results of $\mathcal{O}(\mu)$ double-twist operators and picking these finite spin results up to clarify the points about $\Delta_L$ poles observed in \cite{Fitzpatrick:2019zqz,Fitzpatrick:2019efk,Li:2019tpf,Li:2019zba} up to double-stress-tensor. In fact, the whole of this paper should be viewed as the supplement to \cite{Li:2019zba}: we also study the Regge behavior of lowest-twist double-stress-tensor in the large impact regime and discuss the case in $d=2$, supplementing \cite{Li:2019zba}. In the next subsection, we comment on the mixing problem raised in \cite{Li:2019zba} and argue that heavy-light bootstrap does not suffer from the mixing problem.

\subsection{Comment on the mixing problem}
There are two aspects as referred in \cite{Li:2019zba} that literatures have rarely taken into account. The first one is that the multi-stress-tensor sector has some poles in $\Delta_L$ as observed in \cite{Fitzpatrick:2019zqz,Fitzpatrick:2019efk,Li:2019tpf,Li:2019zba}. As analyzed in \cite{Li:2019zba}, these $\Delta_L$ poles are by-products of multi-stress-tensor trajectories and reflect the fact that multi-stress-tensor $T^k$ could mix with double-trace operators $[\mathcal{O}_L\mathcal{O}_L]$ for those $\Delta_L$ that coincides with the poles. The situation when $\Delta_L$ is specified on the poles is the main topic undertaken in this paper and we leave the details in section \ref{deltapole}.

The second aspect is about the situation that there are additional light operators $\Delta^{(k)}_L=\Delta_L+2k$ where $k\in\mathbb{N}$. In this case, the mixing appears among a family of double-twist operators $[\mathcal{O}_H\mathcal{O}^{(k)}_L]_{n-k,J}$ which shares the same twist and dimension at the leading order. The consequence is that the OPE and anomalous dimension at the order $\mathcal{O}(\mu)$ should be interpreted as the average over the whole family, i.e.
\be
\tilde{c}^{(1)}_{n,J}=\fft{\langle \tilde{c}^{(0)}_{\Delta,J} \tilde{c}^{(1)}_{\Delta,J} \rangle}{\langle \tilde{c}^{(0)}_{\Delta,J} \rangle}\,,\qquad \tilde{\gamma}^{(1)}_{n,J}=\fft{\langle \tilde{c}^{(0)}_{\Delta,J} \tilde{\gamma}^{(1)}_{\Delta,J} \rangle}{\langle \tilde{c}^{(0)}_{\Delta,J} \rangle}\,,\qquad \Delta-J=\Delta_H
+\Delta_L+2n\,.
\ee
The double-stress-tensor at $\mathcal{O}(\mu^2)$ order thus requires us to evaluate $\langle \tilde{c}^{(0)}_{\Delta,J} \big(\tilde{\gamma}^{(1)}_{\Delta,J}\big)^2 \rangle/\langle \tilde{c}^{(0)}_{\Delta,J} \rangle$ instead of simple $\big(\tilde{\gamma}^{(1)}_{n,J}\big)^2$. The similar problem appears in bootstrapping loop corrections of supergravity in AdS$_5\times$S$_5$ \cite{Alday:2017xua}. Nevertheless, in this subsection, we would like to provide an argument to state that there is no mixing in the heavy-light bootstrap.

The key point to resolve the mixing problem, as suggested in \cite{Alday:2017xua}, is to consider the complete set of (mixed) correlators, in our case $\langle\mathcal{O}_H\mathcal{O}^{(p)}_L\mathcal{O}^{(q)}_L\mathcal{O}_H\rangle$. The averaged anomalous dimension now can be organized as a matrix. Let us take a simple example to show this. We consider $n=1$ and there are two light operators $(\Delta_L, \Delta_L^{(1)})$, then the averaged anomalous dimension forms a matrix
\be
\Gamma=\left(
\begin{array}{cc}
 \tilde{\gamma}_{00} & \tilde{\gamma}_{01} \\
 \tilde{\gamma}_{10} & \tilde{\gamma}_{11} \\
\end{array}
\right)\,,
\ee
where $\tilde{\gamma}_{nm}$ stands for the anomalous dimension associated with $\langle\mathcal{O}_H\mathcal{O}^{(m)}_L\mathcal{O}^{(n)}_L\mathcal{O}_H\rangle$ which is contributed by the single stress-tensor sector exchanged in $\langle\mathcal{O}_H\mathcal{O}_H\mathcal{O}^{(m)}_L\mathcal{O}^{(n)}_L\rangle$. The squared anomalous dimension at the next order $\langle \tilde{\gamma}^{(1)2} \rangle$ can thus be read off as the element of the matrix $\Gamma^2$ \cite{Alday:2017xua}. It is now obvious why normally we have $\langle \tilde{\gamma}^{(1)2}\rangle\neq \langle \tilde{\gamma}^{(1)}\rangle^2$. However, in our case, it follows that $\tilde{\gamma}_{01}=\tilde{\gamma}_{10}=0$! This is stemming from a simple fact that the structure constant of scalar-scalar-stress-tensor three-point function $\langle \mathcal{O}_1\mathcal{O}_2T\rangle$ is identically vanishing if the two scalars are different, in other words
\be
\lambda_{\mathcal{O}_1\mathcal{O}_2T}\equiv 0\,,\qquad \text{if $\mathcal{O}_1 \neq \mathcal{O}_2$}.
\ee
One can easily observe this fact from Ward identity associated with stress-tensor which relates $\langle \mathcal{O}_1\mathcal{O}_2T\rangle$ to two-point function $\langle \mathcal{O}_1\mathcal{O}_2\rangle$. As result, the four-point function $\langle\mathcal{O}_H\mathcal{O}_H\mathcal{O}^{(m)}_L\mathcal{O}^{(n)}_L\rangle$ contains no single-stress-tensor and thus gives zero anomalous dimension to its cross-channel double-twist operators. Thus the anomalous dimension matrix in the heavy-light context is automatically diagonal and hence does not admit the mixing.

\section{The Regge pole of double-stress-tensor}
\label{regge}
In this section, we study the Regge behavior of the lowest-twist double-stress-tensor sector, especially in the large impact parameter regime. Normally, the Regge limit of stress-tensors can be used to obtain the cross-channel double-twist anomalous dimensions at large spin limit by means of the impact parameter representation \cite{Kulaxizi:2018dxo}, and the results are consistent with ones extracted from using Lorentzian inversion formula back and forth \cite{Li:2019zba}. Thus it is concluded that we should construct the correct Regge limit from OPE coefficients of lowest-twist multi-stress-tensors. Indeed, it can be observed that the results of lowest-twist double-stress-tensor OPE contain poles in $J$ which are presumably the Regge pole. We would use the Regge pole to find the Regge behavior in an efficient way as \cite{Fitzpatrick:2019efk} did, generalizing what was worked out in $d=4$ \cite{Fitzpatrick:2019efk,Karlsson:2019txu} to general dimensions. However, we can only perform the analysis for double-stress-tensor due to the lack of symbolic $J$ expression for triple and higher stress-tensors. Nevertheless, the Regge behavior for any lowest-twist $k$-stress-tensors in $d=4$ was achieved recently using holography \cite{Karlsson:2019txu}, where the singular behavior turns out to be ${\rm leading}\,\sigma^{-k}+ \text{next-to-leading} \,\sigma^{1-k}$, and we verify this is indeed the case for $k=2$ in general even dimensions.

\subsection{The Regge limit of conformal blocks}
In this subsection, we would like to present the Regge limit of conformal blocks up to next-to-leading singular order in general dimension.

Consider a four-point function decomposed into conformal blocks
\be
\langle \mathcal{O}_1(0)\mathcal{O}_2(z,\bar{z})
\mathcal{O}_3(1)\mathcal{O}_4(\infty)\rangle=\fft{1}{(z\bar{z})^{\fft{\Delta_1+\Delta_2}{2}}}
\sum_{\Delta,J}c_{\Delta,J}G_{\Delta,J}(z,\bar{z})\,,
\ee
the Regge limit is defined as the kinematic limit $z\rightarrow 0, \bar{z}\rightarrow 0$ with fixed $z/\bar{z}$ on the second sheet of $\bar{z}$. To be more precise, we can reach the Regge limit by keeping $z$ fixed from the start, taking $\bar{z}$ to go around $1$ and then sending $z,\bar{z}\rightarrow 0$. Thus we are allowed to go through the following procedure to obtain the Regge conformal block in general dimension order by order:
\begin{itemize}
\item we start with the series expansion of conformal blocks (see Appendix \ref{seriesexp}), pick up the dominant terms in the limit $z\ll \bar{z}$ and then analytically continue $\bar{z}$ by going around $1$, i.e. $(\bar{z}-1)\rightarrow e^{2i\pi}(\bar{z}-1)$ (which will be denoted as $\circlearrowleft$ below), in the end we define $z=\sigma e^\rho, \bar{z}=\sigma e^{-\rho}$ and take $\sigma\rightarrow 0$.
\end{itemize}

As we will see, the Regge behavior of conformal block is singular in $\sigma$ for higher spin $J>2$, and what we are interested is the most two singular terms, i.e. leading and next-to-leading terms. The leading Regge conformal block is well-known as the hyperbolic space propagator \cite{Dolan:2011dv,Kulaxizi:2017ixa} and we would like to follow the mentioned procedure to rederive this. Typically, a certain order of Regge conformal block comes from the same order of series expansion. We thus consider the leading term of series expansion (\ref{seexpand}) in Appdenxi \ref{seriesexp}. By expanding the Gegenbauer polynomial as the series of $z/\bar{z}\rightarrow \infty$ and picking up the dominant terms, we find the Gegenbauer becomes
\be
\tilde{C}_J(\fft{z+\bar{z}}{2\sqrt{z\bar{z}}})\rightarrow \fft{(-1)^J  \Gamma\big(\fft{d-1}{2}\big)\Gamma\big(2J+d-2\big)}
{\Gamma\big(J+d-2\big)\Gamma\big(\fft{1}{2}(2J+d-1)\big)}\xi^J\,_2F_1\Big(\fft{3-d-2J}{2},-J,3-d-2J,\xi^{-1}\Big)\,,
\ee
where $\xi$ is
\be
\xi=\fft{1}{2}(1-\fft{z+\bar{z}}{2\sqrt{z\bar{z}}})\,.
\ee
Going to the second sheet simply flip $(\Delta\rightarrow 1-J, J\rightarrow 1-\Delta)$ with an overall factor $-\fft{i}{\pi}e^{-i\pi(a+b)} (\kappa_{\Delta+J}^{a,b})^{-1}$ \cite{Caron-Huot:2017vep}, we thus find
\bea
i \pi e^{i\pi(a+b)}\kappa_{\Delta+J}^{a,b} G^{(0)\circlearrowleft}_{\Delta,J}&=& \fft{2^{1-d+2\Delta}\sqrt{\pi}\Gamma\big(d-2\Delta\big)}
{\Gamma\big(\fft{1}{2}(1+d-2\Delta)\big)\Gamma\big(\fft{d}{2}-\Delta\big)}\sigma^{1-J} e^{-(1-\Delta)\rho}(1-e^\rho)^{2(1-\Delta)}\times
\cr &&
\cr && \,_2F_1\Big(\Delta-1,\fft{1-d+2\Delta}{2},1-d+2\Delta,-\fft{4e^\rho}{(1-e^\rho)^2}\Big)\,,\label{leadRegge}
\eea
where $G^{(n)}$ is defined by the series expansion of conformal blocks in (\ref{seexpand}). The next-to-leading order is actually very straightforward from the recursion (\ref{recurseCas}) in Appendix \ref{seriesexp}
\be
\sigma^{-1} G^{(1)\circlearrowleft}_{\Delta,J}=-\frac{1}{4}   (\Delta +J-2)G^{(0)\circlearrowleft}_{\Delta,J}\big|_{\Delta\rightarrow \Delta-1}
+\frac{(\Delta -1) (\Delta -d +2) (\Delta -d-J+2)}{(2 \Delta -d +2) (2 \Delta-d )}G^{(0)\circlearrowleft}_{\Delta,J}\big|_{\Delta\rightarrow \Delta+1}\,.\label{nextRegge}
\ee
\subsection{From Sommerfeld-Watson resummation}
\begin{figure}[h]
\centerline{\includegraphics[width=0.45\textwidth]{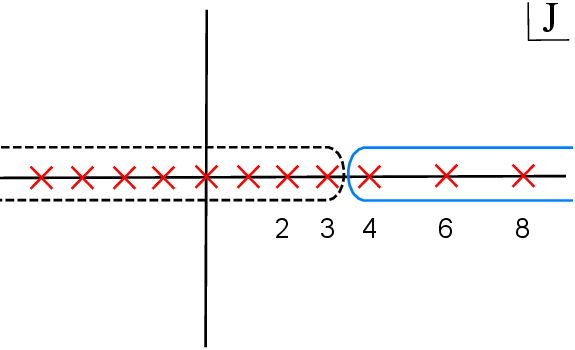}}
\caption{Contour and the deformed contour in the complex $J$ plane.}
\label{Regge}
\end{figure}

With the Regge conformal block in hand, we are now ready to discuss the Regge behavior up to the next-to-leading order of the full lowest-twist double-stress-tensor
\be
\mathcal{G}_{T^2}^{\circlearrowleft}=\sum_{J=4} c_{\Delta,J} G^{\circlearrowleft}_{\Delta,J}\,,\qquad \Delta=2(d-2)\,,\label{Reggesum}
\ee
where the OPE coefficient can be easily computed in even dimensions by following the algorithm proposed in \cite{Li:2019zba}, and the explicit expressions in $d=4,6,8,10$ can be found in \cite{Li:2019zba}, see also \cite{Karlsson:2019dbd} for $d=4,6$. It can be observed that the double-stress-tensor OPE in even dimensions has a Regge pole located at $J=3$ that would allow us to compute (\ref{Reggesum}) without really trying \cite{Fitzpatrick:2019efk}. The trick is to consider the Sommerfeld-Watson transform, writing (\ref{Reggesum}) as
\be
\mathcal{G}_{T^2}^{\circlearrowleft}\sim\int_C dJ \fft{c_{\Delta,J}}{\sin(\pi J)}(1+(-1)^J)
G^{\circlearrowleft}_{\Delta,J}\,,
\ee
where the integration contour $C$ is depicted with blue line in Figure \ref{Regge}. Then we just deform the contour to the opposite direction as the dotted line in Figure \ref{Regge} and pick up those Regge poles $J<4$ encoded in $(1+(-1)^J)c_{\Delta,J}/\sin(\pi J)$.

Since the leading and the next-to-leading Regge behavior goes like $\sigma^{1-J}$ and $\sigma^{2-J}$ respectively, the leading Regge behavior of lowest-twist single-stress-tensor is completely determined by the Regge pole at $J=3$ from the leading Regge conformal block (\ref{leadRegge}) and it behaves as $\sigma^{-2}$. In addition, the next-to-leading Regge behavior is shaped by the Regge pole $J=3$ from the next-to-leading Regge conformal block (\ref{nextRegge}) and $J=2$ from leading one (\ref{leadRegge}).

Although no pattern is found for lowest-twist double-stress-tensor OPE in general even dimensions \cite{Li:2019zba}, we find that the Regge limit of lowest-twist double-stress-tensor sector in general even dimension with large impact parameter limit $\rho\rightarrow\infty$ is generally given by
\bea
&& \mathcal{G}_{T^2}^{\circlearrowleft}=\sigma^{-2} \mathcal{F}_0+ \sigma^{-1}(-\fft{d}{2} e^\rho \mathcal{F}_0+\mathcal{F}_1)\,,
 \cr &&
 \cr && \mathcal{F}_0= -\frac{2^{2d-3} \pi  e^{-2(d-1) \rho } \Gamma (\frac{d+1}{2})^2 \Gamma (-d+\Delta _L+2) \Gamma (\frac{d}{2}+\Delta _L+1)}{\Gamma (1+\frac{d}{2})^2  \Gamma (\Delta _L) \Gamma (-\frac{d}{2}+\Delta _L+1)}\,,
\cr &&
 \cr && \mathcal{F}_1= \frac{-i \pi ^{\fft{1}{2}} 2^{2 d-4} e^{-(2d-3) \rho } \Gamma (\fft{1}{2}+d) \Gamma \left(-d+\Delta _L+2\right) \Gamma \left(\frac{d}{2}+\Delta _L\right)}{\Gamma(d) \Gamma \left(\Delta _L\right) \Gamma \left(-\frac{d}{2}+\Delta _L+1\right)}\,.\label{ReggeTsq}
\eea
Specifying $d=4$, our result (\ref{ReggeTsq}) agrees with the one found in \cite{Karlsson:2019txu,Fitzpatrick:2019efk}.

\section{$[\mathcal{O}_H\mathcal{O}_L]_{n,J}$  with finite spin}
\label{finitej}
In \cite{Li:2019zba}, by using Lorentzian inversion formula, the author shows that the large spin limit of the double-twist operators $[\mathcal{O}_H\mathcal{O}_L]_{n,J}$ at any order of $\mathcal{O}(\mu^k)$ are universal and behave as $J^{-k(d-2)/2}$. In fact, this conclusion can be put forward for leading order $\mathcal{O}(\mu)$: the double-twist operators $[\mathcal{O}_H\mathcal{O}_L]_{n,J}$ at the order $\mathcal{O}(\mu)$ are universal for any $J$. Just notice that if our starting point is the universal single-stress-tensor that is assumed as the only thing contribute to linear $\mathcal{O}(\mu)$ in $[\mathcal{O}_H\mathcal{O}_L]_{n,J}$, then exactly performing the inversion integral can actually lead us to find the universal double-twist $[\mathcal{O}_H\mathcal{O}_L]_{n,J}$ with finite spin support. This conclusion can also be verified from holography as demonstrated in the Appendix \ref{holography}: the higher derivative curvature terms start to show up at $\mathcal{O}(\mu^2)$. As the completion to \cite{Li:2019zba}, we would like to extract the universal $[\mathcal{O}_H\mathcal{O}_L]_{n,J}$ at the order $\mathcal{O}(\mu)$ with finite spin support by using Lorentzian inversion formula in this section.

There is one other thing motivates us to know double-twist operators with finite spin at the linear order. Typically, we wish we can handle the situation when $\Delta_L$ approaches the corresponding poles in lowest-twist multi-stress-tensors, especially for double-stress-tensors. It turns out the pole arises from the integration over large spin \cite{Li:2019zba}, thus the naive expectation is that the summation over finite spin can resolve this problem. We will discuss this shortly in the next section. For this reason, we are more interested in the anomalous dimension rather than OPE, since OPE part will not survive under dDisc \cite{Li:2019zba}.

\subsection{Warm-up: $d=4$}
We first discuss $d=4$ since the full expression of conformal block in $d=4$ is known and rather simple
\bea
&& G^{a,b}_{\Delta,J}(z,\bar{z})=\fft{z \bar{z}}{z-\bar{z}}(k^{a,b}_{\Delta+J}(z)k^{a,b}_{\Delta-J-2}(\bar{z})-
k^{a,b}_{\Delta+J}(\bar{z})k^{a,b}_{\Delta-J-2}(z))\,,
\cr &&
\cr && k^{a,b}_{\beta}(x)=x^{\fft{\beta}{2}}\,_2F_1\Big(a+\fft{\beta}{2},b+\fft{\beta}{2},\beta,x\Big)\,.\label{exactblockd4}
\eea
The conformal block is invariant by interchanging $z$ and $\bar{z}$, thus we can first focus on a simple half of it for simplicity. Then we have the simple half of single-stress-tensor correlator
\bea
\mathcal{G}_T(z,\bar{z})=-\fft{\Delta_L(z-1)^{-1-\Delta_L}(\bar{z}-1)^{1-\Delta_L}(z\bar{z})^{\ft{\Delta_H+\Delta_L}{2}
}\big(3(1-z^2)+(z^2+4z+1)\log z\big)}{40(z-\bar{z})}\,,\label{dimfourT}
\cr &&
\eea
where we slip off the overall $\mu$.

Now there are some differences from large spin limit in \cite{Li:2019zba}. At the large spin limit, one can simply separated (\ref{dimfourT}) into logarithmic part and non-logarithmic part, obtaining anomalous dimension and OPE coefficient respectively. However, this is not true for the data with finite spin, since the logarithmic part can also contribute to OPE coefficient at sub-leading large spin level. In addition, the lightcone expansion of heavy inverted conformal block $G_{J+d-1,\Delta-d+1}$ would be different from \cite{Li:2019zba} due to finite spin effects. Analyzing the Casimir recursion equation (see, e.g. Appendix A in \cite{Caron-Huot:2017vep} and Appendix A in \cite{Li:2019zba}) at the heavy limit but arbitrary spin, we find
\bea
&& \kappa^{a,b}_\beta\mu^{a,b}(z,\bar{z})G_{J+d-1,\Delta-d+1}(z,\bar{z})=\sum_{p=0}\tilde{B}^{a,b}_{p}\kappa^{a,b}_{\beta+2p}
z^{\fft{J-\Delta}{2}+p-1}k^{a,b}_{\beta+2p}(\bar{z})\,,
\cr &&
\cr && \tilde{B}^{a,b}_{p}=\frac{(-1)^p \Gamma \left(\frac{d}{2}\right) (J+2 p) \Gamma \left(\frac{d}{2}+J\right) \Gamma (J+p)}{\Gamma (J+1) \Gamma (p+1) \Gamma \left(\frac{d}{2}-p\right) \Gamma \left(\frac{d}{2}+J+p\right)}\,,\label{expandheavyfinj}
\eea
where $\kappa^{a,b}_\beta$ and $\mu^{a,b}(z,\bar z)$ can be found in Appendix \ref{inversion} and $a=b=1/2(\Delta_L-\Delta_H)$. Thus the inversion formula (\ref{Loreninv}) becomes
\be
\tilde{c}(n',J)=\fft{1}{2}\int dzd\bar{z}\sum_{p=0}\tilde{B}^{a,b}_{p}\kappa^{a,b}_{\beta+2p}
z^{\fft{J-\Delta}{2}+p-1}k^{a,b}_{\beta+2p}(\bar{z}){\rm dDisc}[\mathcal{G}_T]\,,\label{intfinJ}
\ee
where $n'$ is defined by $\Delta-J=\Delta_H+\Delta_L+2n'$. To perform above inversion integral, we expand $\mathcal{G}_T$ in terms of $z$ and $(1-\bar{z})/\bar{z}$ and then use the formula (\ref{intresult}) termwise, in the end, we just need to work out the remaining integral over $z$. We still work on logarithmic part and non-logarithmic part separately just to see clearly how $\log z$ contributes to OPE coefficient. With the unit of free OPE coefficient at the heavy limit (\ref{freeOPEheav}), we find the non-logarithmic part gives us a rather simple expression
\be
\tilde{c}^{(1){\rm non-log}}_{n,J}=-\fft{3\Delta_L(\Delta_L+2n-1)}{4(J+1)}\,.\label{fourdimcp1}
\ee
The logarithmic part is a bit complicated, typically, the integration with $\log z$ will provide us the factor
\be
\fft{1}{\big(\Delta-J-(\Delta_H+\Delta_L+2n)\big)^2}\,.
\ee
We parameterize $\Delta-J=\Delta_H+\Delta_L+2n'$ and then the inversion integral is evaluated to give us
\bea
\tilde{c}(n',J)=\frac{ \left(\Delta _L^2-\Delta _L+6 n \Delta _L+6 n^2-6 n\right) \Gamma (J+n+2)\Gamma \left(J+n'+\Delta _L\right)}{4 (J+1)  \Gamma (J+n'+2) \Gamma \left(J+n+\Delta _L\right)(n-n')^2}\,.\label{dim4data}
\eea
Now it is clear that why logarithmic part would also contribute a part of OPE coefficient: the OPE data (\ref{dim4data}) contains $n'$ in the numerator, then evaluating the OPE data around the physical twist $n$ would result in additional $n'-n$ and thus give us a part of OPE coefficient. Specifically, by ignoring the regular terms we have
\be
\tilde{c}(n',J)=\frac{ \left(\Delta _L^2-\Delta _L+6 n \Delta _L+6 n^2-6 n\right) }{4 (J+1) (n-n')^2}\Big(1+(\psi_{n+J+2}-\psi_{n+J+\Delta_L})(n-n')\Big)\,.
\ee
Thus the OPE and anomalous dimensions can be readily read off from the pole structures
\bea
\cr && \tilde{\gamma}^{(1)}_{n,J}=-\frac{ \left(\Delta _L^2-\Delta _L+6 n \Delta _L+6 n^2-6 n\right) }{2(J+1)}\,,\qquad \tilde{c}^{(1){\rm log}}_{n,J}=
-\fft{1}{2}\tilde{\gamma}^{(1)}_{n,J}(\psi_{n+J+2}-\psi_{n+J+\Delta_L})\,.\label{dim4gaope}
\cr &&
\eea

We can then evaluate the OPE data from another half of stress-tensor block $\mathcal{G}_T(\bar{z},z)$ by following above calculations. Since there is no $\log z$ in $\mathcal{G}_T(\bar{z},z)$, it only gives rise to single pole of $n^\prime-n$ and thus does not contribute to anomalous dimensions. However, it nontrivially modifies the OPE coefficient at finite spin by inverting $\log\bar{z}$. Again by expanding in terms of $z$ and $(1-\bar{z})/\bar{z}$ we find appended contributions
\bea
\tilde{c}_{n,J}^{(1){\rm app}}=\fft{\Delta_L-2}{2\Delta_L}(3\Delta_L-2\tilde{c}^{(1){\rm non-log}}_{n,J}) + \tilde{c}_{n,J}^{(1){\rm log}} \big(1-\fft{3(\Delta_L+2n+J)}{\tilde{\gamma}_{n,J}^{(1)}}\big)\,.
\eea

In summary, in $d=4$, we find the anomalous dimension of double-twist operators $[\mathcal{O}_H\mathcal{O}_L]_{n,J}$ with finite spin is given by the first formula in (\ref{dim4gaope}) and the OPE coefficient is given by \footnote{The contribution $\tilde{c}_{n,J}^{(1){\rm app}}$ was missing in previous version of this paper by mistake. We are grateful to Alba Grassi, Cristoforo Iossa, Daniel Panea Lichting, Matthew Dodelson and Sasha Zhiboedov for pointing this out to us.}
\be
\tilde{c}^{(1)}_{n,J}=\tilde{c}^{(1){\rm non-log}}_{n,J}+\tilde{c}^{(1){\rm log}}_{n,J}+\tilde{c}_{n,J}^{(1){\rm app}}\,,
\ee
where the specific details can be found in (\ref{fourdimcp1}) and the second formula in (\ref{dim4gaope}). Whenever our results are expanded in terms of $1/J$, the agreements with \cite{Karlsson:2020ghx} can be observed.

We should state that our results are valid to any $J$, including $J=0$. Standardly, Lorentzian inversion formula may not applicable for $J<2$ \cite{Caron-Huot:2017vep,Simmons-Duffin:2017nub}. A simple and intuitive reason is that normally the inversion integral is not convergent in the regime $J<2$ \cite{Simmons-Duffin:2017nub}, such that the resulting OPE or anomalous dimension poses some poles in $J$ within the region $0\leq J <2$. Fortunately, we can see that our results with finite spin at the heavy-limit (\ref{fourdimcp1}) and (\ref{dim4gaope}) is free of any non-negative poles in $J$. In fact, we will see shortly that the anomalous dimensions with finite spin are well-behaved in all $J\geq 0$ in any dimensions. However, even though the OPE coefficients and anomalous dimensions do not have any poles in $J\geq0$, it is still possible that the result of $J=0$ is not reliable due to the bad Regge behavior of the correlator $\mathcal{G}$ in the dDisc: it stays bounded rather than decays fast enough such that the arcs at infinity cannot be dropped in the derivation of the Lorentzian inversion formula \cite{Caron-Huot:2017vep,Simmons-Duffin:2017nub}. In this way, the subtraction is required such that the Regge behavior is made nicer \cite{Carmi:2019cub}, which would cause the ambiguity of the correlator associated with spin zero. More precisely, by perturbing the OPE and anomalous dimensions with $J=0$ appropriately, the correlator can be enhanced to have additional crossing-symmetric part that vanishes under dDisc, e.g. see \cite{Alday:2017vkk}, thus the OPE and anomalous dimensions at $J=0$ obtained from the Lorentzian inversion formula are far from complete. Nevertheless, the heavy-limit we consider removes all the subtleties. It is worth noting the correlator $\mathcal{G}$ in our case has an overall $(z\bar{z})^{(\Delta_H+\Delta_L)/2}$ that highly suppresses the Regge behavior of the wanted correlator by $z\bar{z}\rightarrow0, \Delta_H\rightarrow \infty$. Another justification is the heavy block that we would present in the next section (\ref{gdimheav}), it is obvious that there is no way to construct additional crossing-symmetric correlator associated with $J=0$. Moreover, we can also reproduce the anomalous dimension in (\ref{dim4gaope}) from holography by using the Hamiltonian perturbation approach developed in \cite{Fitzpatrick:2014vua}, see Appendix \ref{holography}. Since there is no similar subtlety for $J=0$ has been observed in holographic theory, we thus verify that our results with $J=0$ should hold true. Hence, we should trust that they are analytic in all non-negative spin. This argument is essential, since we are going to sum over conformal blocks from $J=0$ in the next section.

\subsection{General dimensions}
The obstacle of extracting the double-twist $[\mathcal{O}_H\mathcal{O}_L]_{n,J}$ with finite spin is that our knowledge of $\bar{z}\rightarrow 1$ expansion of the single-stress-tensor conformal block in general dimension is limited. Recently, \cite{Li:2019dix,Li:2019cwm} have obtained the $\bar{z}\rightarrow 1$ expansion of conformal blocks with any $(\Delta,J,a,b)$. For our purpose, we need $(\Delta=d, J=2)$ and $a=b=0$. However, it would be more convenient to have single-stress-tensor block expanded in term of $(1-\bar{z})/\bar{z}$ rather than $(1-\bar{z})$ as in \cite{Li:2019dix,Li:2019cwm}. In addition, as we mentioned above, we are only interested in the anomalous dimension in general dimensions, thus the logarithmic part of $(1-\bar{z})/\bar{z}$ expansion of stress-tensor block is enough for our purpose. From the geodesic Witten diagram \cite{Hijano:2015zsa}, we manage to find a nice expression for $(1-\bar{z})/\bar{z}$ expansion of stress-tensor conformal block with parameterizing $y=z/(1-z), \bar{y}=(1-\bar{z})/\bar{z}$
\bea
&& G_{T}(y,\bar{y})=\sum_{k=0} \mathcal{N}_k\fft{\bar{y}^k\,y^{\fft{d-2}{2}}\log \bar{y} }{y+1}\big((d-2) (3 d (y+1)+2 (k y+k-2 y-1)) g_{\fft{d}{2},k}(y)
\cr && -2 \big(2 d^2 (y+1)+d (k (4 y+3)-6 y-5)+2 (k-1) (k y+k-2 y-1)\big) g_{\fft{d-2}{2},k}(y)\big)\,,
\cr &&\label{stresszb11}
\eea
where the coefficient $\mathcal{N}_k$ and the function $g_{p,k}(y)$ is given below
\bea
&& \mathcal{N}_k= \frac{2^{d+1} \Gamma \left(\frac{d+3}{2}\right)  \Gamma \left(\frac{d}{2}+k+1\right)}{\sqrt{\pi }  (d+2 k-2) (d+2 k) \Gamma \left(\frac{d}{2}+1\right) \Gamma (k+1)^2 \Gamma \left(\frac{d}{2}-k+1\right)}\,,
\cr &&
\cr && g_{p,k}(y)=\,_2F_1\big(p,-k,\ft{1}{2}(d+2-2k),-y\big)\,.\label{stresszb12}
\eea
As far as we know, (\ref{stresszb11}) is a new result, and the trick of this guessing job is listed in Appendix \ref{geowitten}. We verify that our results (\ref{stresszb11}) are consistent with the general $\bar{z}\rightarrow 1$ expansion of conformal block found in \cite{Li:2019dix,Li:2019cwm} up to very high order of $\bar{y}$. Taking $d=4$, we can easily verify that (\ref{stresszb11}) reproduces the correct logarithmic part. Then we have
\be
\mathcal{G}_{T}(z,\bar{z})=\fft{(z\bar{z})^{\fft{\Delta_H+\Delta_L}{2}}}{((1-z)(1-\bar{z}))^{\Delta_L}}G_{T}(\bar{y},y)\,.
\ee
We can then apply the formula (\ref{intresult}) to (\ref{intfinJ}) termwise, leading to a final integral with (\ref{expandheavyfinj}) to be integrated over $z$. The remaining integral is not possible to be done once and for all, we are supposed to expand the integrand as a series of $z$ and extract the OPE coefficients twist by twist. Surprisingly, we find that the results show a pattern: there is a series representation for the anomalous dimension with finite spin!
\bea
\tilde{\gamma}^{(1)}_{n,J}=\sum_{p=0}^n \sum_{q=0}^{n-p} \ft{(-1)^{p+1}\Gamma(J+1)\Gamma(q+\fft{d+2}{2})\Gamma(p+\fft{d-2}{2})\Gamma(p+q-n)\Gamma(\fft{d}{2}-q-\Delta_L)\Gamma(\Delta_L+1)\sin(\fft{\pi}{2}
(d-2\Delta_L))}{2\pi \Gamma(q+1)\Gamma(p+1)\Gamma(\fft{d+2}{2}-q)\Gamma(\fft{d-2}{2}-p)\Gamma(\fft{d}{2}+J+p)\Gamma(1+p+q)\Gamma(-n)}\,.
\cr &&\label{anogenedim}
\eea
It can be verified that for $d=4$ (\ref{anogenedim}) comes back to the first formula in (\ref{dim4gaope}). By the same argument, (\ref{anogenedim}) is expected to be valid for $J=0$. The large $n$ limit looks nice and is given by \footnote{Note this behavior is different from \cite{Kulaxizi:2018dxo} where it goes $n^{d/2}$. The reason is that the limit $J\gg n\gg 1$ is taken in \cite{Kulaxizi:2018dxo} while we keep $J$ finite.}
\be
\tilde{\gamma}^{(1)}_{n,J}|_{n\rightarrow \infty}= \fft{(d-1)\Gamma(J+1)n^{d-2}}{\Gamma(J+d-2)}\,.\label{largen}
\ee
Normally, for all light operators, the large $n$ limit OPE and anomalous dimension is dominated near the bulk-point singularity $z-\bar{z}\rightarrow 0$ (on the second sheet) where the four-point function can be mapped to the massless four-point amplitude in flat-space \cite{Gary:2009ae,Okuda:2010ym,Penedones:2010ue,Maldacena:2015iua,Heemskerk:2009pn}, for which the large $n$ limit of anomalous dimension shall predict the coefficients of partial wave expansion of the massless amplitude with noting $n^2\sim s$ \cite{Alday:2017vkk}. However, the heavy-limit where $\Delta_H\gg n$ seems to ruin every nice set-ups of the standard flat space limit. What is the flat-space amplitude the large $n$ limit of heavy-light four-point function (\ref{largen}) corresponds to is thus a tantalizing question to ask in the future.

\subsubsection{Explicit examples in $d=6,8,10$}

For latter uses, we would like to present the explicit examples in $d=6,8,10$.

In $d=6$, we find
\bea
&& \tilde{\gamma}^{(1)}_{n,J}=-\fft{1}{2 (J+1) (J+2) (J+3)}\big(3 \left(J (4 n-1)+4 n^2+6 n-3\right) \Delta _L^2+(J+2 n+3) \Delta _L^3
\cr && +2 (3 J n (5 n-7)+J+n (n (10 n+9)-35)+3) \Delta _L+10 (n-2) (n-1) n (2 J+n+3)\big)\,,
\cr &&\label{d6}
\eea
In $d=8$ we have
\bea
&& \tilde{\gamma}^{(1)}_{n,J}=-\fft{1}{2 (J+1) (J+2) (J+3)(J+4)(J+5)}\big(2 n^2 (5 J^2 (9 \Delta _L^2-51 \Delta _L+77)
\cr && +15 J (\Delta _L (\Delta _L+6) (2 \Delta _L-9)+91)+3 \Delta _L (\Delta _L (\Delta _L (\Delta _L+24)-64)-136)+1036)
\cr && +2 n (5 J^2 (\Delta _L-3) (\Delta _L (2 \Delta _L-9)+14)+3 J (\Delta _L (\Delta _L (\Delta _L (\Delta _L+14)-124)+319)-294)
\cr && +2 \Delta _L (3 \Delta _L (\Delta _L (2 \Delta _L+3)-68)+518)-840)+(J+4) (J+5) (\Delta _L-3) (\Delta _L-2) (\Delta _L-1) \Delta _L
\cr && +84 n^5 (J+\Delta _L)+10 n^4 (3 (7 J+4) \Delta _L+7 (J-3) J+9 \Delta _L^2-42)+20 n^3 (J+\Delta _L) (7 J (\Delta _L-3)
\cr && +2 \Delta _L (\Delta _L+6)-42)+28 n^6\big)\,.\label{d8}
\eea
The case $d=10$ is highly complicated and is given by
\bea
&& \tilde{\gamma}^{(1)}_{n,J}=-\fft{1}{2 (J+1) (J+2) (J+3)(J+4)(J+5)(J+6)(J+7)}\big(28 n^6 (18 J^2+9 J (5 \Delta _L-4)
\cr && +5 \Delta _L (4 \Delta _L+5)-99)+84 n^5 (J+\Delta _L) (3 J^2+3 J (5 \Delta _L-12)+5 \Delta _L (\Delta _L+5)-99)
\cr && +20 n^3 (J+\Delta _L) (7 J^2 (\Delta _L (4 \Delta _L-31)+63)+7 J (\Delta _L (\Delta _L (2 \Delta _L+13)-159)+324)
\cr && +\Delta _L (\Delta _L+5) (\Delta _L (\Delta _L+45)-256)+2709)+2 n ((6 J^2+63 J+167) \Delta _L^5+5 (3 J (J+1) (J+7)
\cr && -130) \Delta _L^4-5 (3 J (J (13 J+128)+378)+811) \Delta _L^3+10 (J (J (104 J+987)+2989)+2765) \Delta _L^2
\cr && -2 (J (J (1375 J+11976)+33173)+28782) \Delta _L+432 (J (7 J (J+8)+139)+105))
\cr && +10 n^4 (63 J^3 (\Delta _L-4)+42 J^2 (\Delta _L (4 \Delta _L-13)-6)+21 J (\Delta _L (\Delta _L (5 \Delta _L+9)-119)+108)
\cr && +\Delta _L (\Delta _L (15 \Delta _L (\Delta _L+15)-724)-1280)+2709)+2 n^2 (105 J^3 (\Delta _L-4) ((\Delta _L-7) \Delta _L+15)
\cr && +6 J^2 (5 \Delta _L (\Delta _L (3 \Delta _L (\Delta _L+1)-226)+1039)-7392)+3 J (5 \Delta _L (\Delta _L (\Delta _L (\Delta _L (\Delta _L+38)-211)
\cr && -432)+4069)-31332)+5 \Delta _L (\Delta _L (3 \Delta _L (\Delta _L (5 \Delta _L+47)-470)+1898)+5530)-57564)
\cr && +(J+5) (J+6) (J+7) (\Delta _L-4) (\Delta _L-3) (\Delta _L-2) (\Delta _L-1) \Delta _L+360 n^7 (J+\Delta _L)+90 n^8\big)\,.\label{d10}
\cr &&
\eea
We can also obtain these explicit examples by using the holographic Hamiltonian perturbation approach \cite{Fitzpatrick:2014vua} in Appendix \ref{holography}.
\section{On the $\Delta_L$ poles}
\label{deltapole}
It turns out that in general the multi-stress-tensor OPE coefficients contain some poles in $\Delta_L$  \cite{Fitzpatrick:2019zqz,Fitzpatrick:2019efk,Li:2019tpf,Li:2019zba}. It is argued that when $\Delta_L$ takes the value of those poles, the multi-stress-tensor sectors $T^k$ mix with double-trace sectors $[\mathcal{O}_L\mathcal{O}_L]_{n,J}$ such that the complete correlator is free of divergence \cite{Fitzpatrick:2019zqz} with the mixed OPE denoted as $c^{\rm mix}$. In addition, the mixed operators acquire anomalous dimensions $\gamma^{\rm mix}$, and it turns out that the product of these anomalous dimensions and the relevant OPE coefficients obey the so called Residue relation to the corresponding multi-stress-tensor OPE coefficients $c_{T^k}$ \cite{Li:2019tpf}: taking the Residue of the relevant multi-stress-tensor OPE coefficients at the poles correctly give rise to the product of the mixed OPE coefficient $c^{\rm mix}$ and anomalous dimension $\gamma^{\rm mix}$, i.e.,
\be
\gamma^{\rm mix}c^{\rm mix}=-2{\rm Res}_{\Delta_L={\rm poles}}c_{T^k}\,.\label{rrl}
\ee
However, the mixed OPE $c^{\rm mix}$ is beyond our knowledge even in the holographic context  \cite{Fitzpatrick:2019zqz,Li:2019tpf}.

In this section, we aim to discuss the situation when $\Delta_L$ approaches the poles of multi-stress-tensor OPE. Our main focus is $d=4$ where we take $\Delta_L=2$, we extract the mixed OPE coefficient and verify the Residue relation. For higher dimensions, we take $d=6$ as an example to verify the Residue relation. With the anomalous dimensions (\ref{d8}) and (\ref{d10}) in hands, we can easily verify Residue relation in $d=8,10$, but we will not provide the details of $d=10$ in the paper, for $d=8$ see Appendix \ref{mixOPE}. We leave the examples of the OPE coefficients in $d=6,8$ in Appendix \ref{mixOPE}. In the end, we understand the universality of the Residue relation.

\subsection{Summation instead of integration}
Typically, the $\Delta_L$ poles arise from the integration over $J$ in the large spin limit \cite{Li:2019zba}. It turns out that it boils down to following integral
\be
\int_0^{\infty} dJ \fft{\bar{z}^J}{J^a}=(-1)^{a-1}\Gamma(1-a)(\log \bar{z})^{a-1}\,.\label{toyint}
\ee
We thus see the poles $\Gamma(1-a)$ appear and whenever we take, for example, $a=1$, the integral diverges. This divergence actually comes from the ill-behaved integrand at $J=0$. To be clear, we can introduce the cut-off $\epsilon\rightarrow 0$ and then we have
\be
\int_\epsilon dJ \fft{\bar{z}^J}{J}\sim \log \epsilon\,.
\ee
How do we resolve this thing? Since the large spin limit of the anomalous dimensions is ill-behaved, the most natural consideration is then to consider the anomalous dimension with finite spin that have been worked out in the previous section. It is apparent from the last section that the anomalous dimensions with finite spin are well-behaved at $J=0$. Since there is no any large spin limit presumed, we are required to sum over $J$ rather than integrate over it in the procedure of obtaining the correlators. We would like to present the specific procedure here.

Like \cite{Li:2019zba}, we need the heavy block as the ingredient for us to perform the summation in order to obtain the required correlators. In $d=4$, we just need to take the heavy limit of a half of the full conformal block (\ref{exactblockd4}), resulting in
\be
g_{n,J}(z,\bar{z})=\fft{(z\bar{z})^{n+\fft{\Delta_H+\Delta_L+\tilde{\gamma}}{2}}}{\bar{z}-z}\bar{z}^{J+1}\,,\label{g4heav}
\ee
where $\tilde{\gamma}$ is the anomalous dimensions for double-twist operators, and $g_{n,J}$ denotes the conformal block $G_{\Delta,J}^{a,b}$ at the heavy-limit that specifies $\Delta-J=\Delta_H+\Delta_L+2n+\tilde{\gamma}$. In general dimension, we have to solve the Casimir equation for lightcone expansion of conformal blocks with heavy limit as in \cite{Li:2019zba}, i.e. solve $B^{a,b}_{pq}$ in
\be
G^{a,b}_{\Delta,J}(z,\bar{z})=\sum_{p}\sum_{q=-p}^{q=p} B^{a,b}_{pq} z^{\fft{1}{2}(2(n+p) +\Delta_L+\Delta_H+\tilde{\gamma})}\bar{z}^{\ft{\Delta_H+\Delta_L+\tilde{\gamma}}{2}+J+q+n}
\,.\label{HLLH-hevex}
\ee
The difference from \cite{Li:2019zba} is that \cite{Li:2019zba} takes the large spin limit in addition to heavy-limit. Following Appendix A in \cite{Li:2019zba} but keep $J$ finite, we find
\be
B^{a,b}_{pq}=\frac{\Gamma (J+1) \Gamma \left(\frac{1}{2} (d+2 p-2)\right) \Gamma \left(\frac{1}{2} (d+2 J-2 p-2)\right)}{p! \Gamma \left(\frac{d-2}{2}\right) \Gamma \left(\frac{1}{2} (d+2 J-2)\right) \Gamma (J-p+1)}\delta_{p+q,0}\,,
\ee
which leads us to
\be
g_{n,J}(z,\bar{z})=z^{\fft{\Delta_H+\Delta_L+\tilde{\gamma}}{2}+n}\bar{z}^{\fft{\Delta_H+\Delta_L+\tilde{\gamma}}{2}+n+J}
\,_2F_1\big(\fft{d-2}{2},-J,2-\fft{d}{2}-J,\fft{z}{\bar{z}}\big)\,.
\label{gdimheav}
\ee
Then we shall evaluate the part of correlator that remains under dDisc
\be
\sum_{n=0,J=0}\tilde{c}^{\rm free}_{n,J} \fft{\tilde{\gamma}^{(1)2}_{n,J}}{8}(\log z)^2 g_{n,J}(z,\bar{z})
\big|_{\Delta_L\rightarrow{\rm poles}}\,.\label{squreansa}
\ee
Crossing it to $(z\rightarrow 1-\bar{z},\bar{z}\rightarrow 1-z)$ and keeping terms up to one that produces the lowest-twist double-stress-tensor, the resulting correlator should give us the mixed OPE coefficients and anomalous dimensions. We only consider lowest-twist double-stress-tensor, because this is the only thing we can trust: only the lowest-twist multi-stress-tensor is universal \cite{Li:2019zba}. For higher twist cases, there will be non-universal piece beyond our awareness engaging in the game \cite{Karlsson:2020ghx} such that the mixed results would be unwarranted. The exception is CFT$_2$ where there is no higher twist double-stress-tensor exists at all! This will be discussed momentarily in section \ref{CFTtwo}. In general, as we will see in $d=6$ and $d=8$ in Appendix \ref{mixOPE}, the lowest-twist double-stress-tensor may not be leading-twist double-trace operator $[\mathcal{O}_L\mathcal{O}_L]_{n,J}$, in other words, it may mix with double-trace operators with nonzero $n=n_0>0$. At the order $\mathcal{O}(\mu^2)$, those double-trace operators are also all universal.

\subsection{Verify the Residue relation}
\subsubsection{$d=4$}
In $d=4$, the only pole is $\Delta_L=2$, for which the lowest-twist double-stress-tensor mixes with lowest-twist of double-trace operator $[\mathcal{O}_L\mathcal{O}_L]_{0,J}$. The free OPE and anomalous dimensions simplify a lot with $\Delta_L=2$
\be
\tilde{c}^{\rm free}_{n,J}=1+J\,,\qquad \tilde{\gamma}^{(1)}_{n,J}=-\fft{1+3n+3n^2}{1+J}\,.
\ee
Evaluating from (\ref{squreansa}), we actually find a rather simple crossed correlator
\be
\mathcal{G}(z,\bar{z})\big|_{\log(1-\bar{z})^2}=-z^2 \log(\fft{z}{2})\,\fft{216-432\bar{z}+264 \bar{z}^2-48\bar{z}^3+\bar{z}^4}{16\bar{z}^4}\,.
\ee
Then we just need to work on the inversion integral, restricting to the lowest-twist, (\ref{Loreninv}) becomes
\be
\kappa_\beta \int dzd\bar{z}\mu^{0,0}(z,\bar{z})z^{\fft{2-\tau}{2}}k_\beta^{0,0} {\rm dDisc}[\mathcal{G}(z,\bar{z})\big|_{\log(1-\bar{z})^2}]\,.
\label{intsimdim4}
\ee
Following integral formula would be useful
\be
\int \fft{d\bar{z}}{\bar{z}^2} k^{0,0}_\beta(\bar{z})\big(\fft{1-\bar{z}}{\bar{z}}\big)^a=-\fft{2\Gamma(a+1)^2 \Gamma(\fft{\beta}{2}-a)\Gamma(\beta)}
{(2(a+1)-\beta)\Gamma(a+1+\fft{\beta}{2})\Gamma(\fft{\beta}{2})^2}\,.
\ee
We thus find
\bea
&& c(\beta,J)|_{n\rightarrow0}=(\fft{1}{2n^2}-\fft{\log 2}{2n})\eta(\beta)\,,\qquad \eta(\beta)=\frac{\sqrt{\pi } 2^{4-\beta } \left(\beta ^4-4 \beta ^3+92 \beta ^2-176 \beta +384\right) \Gamma \left(\frac{\beta-2 }{2}\right)}{(\beta -10) (\beta -6) \beta  (\beta +4) (\beta +8) \Gamma \left(\frac{\beta-1 }{2}\right)}\,,\label{mixdatadim4}
\cr &&
\eea
where $\beta=4+2J+2n$. From (\ref{relacandano}), the coefficient of $1/n^2$ immediately gives rise to
\bea
\gamma^{\rm mix} c^{\rm mix}=-2\eta|_{\beta=4+2J}=-\frac{\sqrt{\pi } 2^{-2 J-2} J \left(J^4+6 J^3+35 J^2+78 J+72\right) \Gamma (J-1)}{(J-3) (J+2) (J+4) (J+6) \Gamma \left(J+\frac{3}{2}\right)}\,.
\cr &&
\eea
Having the explicit lowest-twist double-stress-tensor OPE $c_{T^2}$ \cite{Fitzpatrick:2019efk,Kulaxizi:2019tkd,Li:2019zba}, we can then verify the Residue relation proposed in \cite{Li:2019tpf}
\be
\gamma^{\rm mix} c^{\rm mix}=-2{\rm Res}_{\Delta_L=2}\,c_{T^2}\,.
\ee
Furthermore, we can also extract the mixed OPE coefficient from (\ref{mixdatadim4})
\be
c^{\rm mix}=(-2\partial_\beta+\log 2)\eta(\beta)\big|_{\beta=4+2J}\,.
\ee

\subsubsection{$d=6$}
In $d=6$, from the results in \cite{Li:2019zba,Karlsson:2019dbd}, there are two poles in $\Delta_L$: $\Delta_L=4$ and $\Delta_L=3$. For $\Delta_L=4$, the lowest-twist double-stress-tensor mixes with the leading-twist double-trace operator $[\mathcal{O}_L\mathcal{O}_L]_{0,J}$; for $\Delta_L=3$, the lowest-twist double-stress-tensor mixes with the sub-leading-twist double-trace operator $[\mathcal{O}_L\mathcal{O}_L]_{1,J}$, which will be naturally observed shortly.

For $\Delta_L=4$, the summed correlator $\mathcal{G}$ is cumbersome, and we are not going to present the complete expression here but just exhibit the logarithmic part that looks much simpler
\be
\mathcal{G}(z,\bar{z})|_{\log(1-\bar{z})^2, \log z}\sim -\frac{3 z^4 \left(\bar{z}^6-96 \bar{z}^5+1296 \bar{z}^4-5900 \bar{z}^3+11700 \bar{z}^2-10500 \bar{z}+3500\right) }{8 \bar{z}^6}\,.
\ee
Then the inversion integral like (\ref{intsimdim4}) (replace $2-\tau$ by $4-\tau$) would give us
\be
\gamma^{\rm mix}c^{\rm mix}=-\frac{3 \sqrt{\pi } 2^{-2 J-3} \left(J^4+14 J^3+101 J^2+364 J+540\right) \Gamma (J+4)}{(J-3) (J-1) (J+1) (J+6) (J+8) (J+10) \Gamma \left(J+\frac{7}{2}\right)}\,.
\ee
From lowest-twist double-stress-tensor OPE in $d=6$ obtained in \cite{Li:2019zba,Karlsson:2019dbd}, we can verify the Residue relation
\be
\gamma^{\rm mix} c^{\rm mix}=-2{\rm Res}_{\Delta_L=4}\,c_{T^2}\,.
\ee
By including the non-logarithmic part, we can extract the mixed OPE $c^{\rm mix}$ which is presented in the Appendix \ref{mixOPE}.

For $\Delta_L=3$, the summed correlator is still too complicated to be presented here, but we would like to comment the feature we observe. We find the correlator behaves as follows
\be
\mathcal{G}(z,\bar{z})=z^3 f_1(\bar{z})+ z^4 f_2(\bar{z})+ z^4 \log z\,f_3(\bar{z})\,,
\ee
where $f_i(\bar{z})$ represents something depending on $\bar{z}$. The logarithmic part appears with higher order in $z$! In fact, the power $z^3$ gives rise to the operator with twist given by $\Delta-J=6$. It is not possible for stress-tensors to have this twist, the only possibility is the leading-twist double-trace operator $[\mathcal{O}_L\mathcal{O}_L]_{0,J}$ which is free of mixing, implied by that it contains no $\log z$. The power $z^4$ would lead to the trajectory with $\Delta-J=8$, this is a mixture from lowest-twist double-stress-tensor and sub-leading-twist double-trace operator $[\mathcal{O}_L\mathcal{O}_L]_{1,J}$. $f_3(\bar{z})$ is rather simple
\be
f_3(\bar{z})\big|_{\log(1-\bar{z})^2}=\frac{9 \left(\bar{z}^9-168 \bar{z}^8+2688 \bar{z}^7-13040 \bar{z}^6+26520 \bar{z}^5-24000 \bar{z}^4+8000 \bar{z}^3\right) }{128 \bar{z}^9}\,,
\ee
Then we can evaluate the Lorentzian inversion formula to obtain
\be
\gamma^{\rm mix}c^{\rm mix}=\frac{9 \sqrt{\pi } 2^{-2 J-7} \left(J^6+21 J^5+283 J^4+2247 J^3+10372 J^2+25956 J+27360\right) \Gamma (J+3)}{(J-3) (J-1) (J+1) (J+4) (J+6) (J+8) (J+10) \Gamma \left(J+\frac{7}{2}\right)}\,,\label{canodim6}
\ee
and then verify
\be
\gamma^{\rm mix} c^{\rm mix}=-2{\rm Res}_{\Delta_L=3}\,c_{T^2}\,.
\ee
We leave the leading-twist double-trace OPE coefficient and the mixed OPE coefficients in Appendix \ref{mixOPE}.

We actually verified the Residue relation for $d=8$ and $d=10$ by using the obtained double-twist anomalous dimension with finite spin (\ref{d8}) and (\ref{d10}). As in $d=6$, the common property is that the maximal pole corresponds to the mixing with leading-twist double-trace operators, and the next pole corresponds to sub-leading-twist and so on.

\subsection{The universality of the Residue relation}
In this subsection, we would like to try to understand the universality of the Residue relation. Since the lowest-twist double-stress-tensor is universally fixed by large spin region of its cross channel, the Residue relation immediately implies that for $\Delta_L$ located at poles, the logarithmic part of evaluated correlator is also completely controlled by the large spin kinematics of cross channel. To see this fact, let us think what might be the origin of the logarithmic behavior?

As quoted in \cite{Li:2019zba}, we can simply take $\Delta_L$ approach a certain pole, then the large spin correlator would create a $\log z$ term in addition to the divergent part (let us introduce a cut-off to regulate it), with coefficients given by Residue of double-stress-tensor OPE coefficient. However, naively doing this bothers us with the divergent part: we should not have divergence at all with getting the correct logarithmic part. To resolve the divergence, we shall recall that the divergence comes from $J=0$. However, remember we are working in the large spin limit, implying $J=0$ is a fake value where we integrate from. We can simply go from $J=1$ instead, then there will be no any divergent term but logarithmic term kept same. To be precise, we look at the integral
\be
\int_1^\infty dJ \fft{\bar{z}^J}{J^a}=\int_\epsilon^\infty dJ \fft{\bar{z}^J}{J^a}- \int_\epsilon^1 dJ \fft{\bar{z}^J}{J^a}\,,\label{intfromone}
\ee
where $\epsilon\rightarrow 0$ is the cut-off. Sending $a$ to the poles, the logarithmic term $\log z$ after crossing only appears from the first term in (\ref{intfromone}) with the divergence canceled identically by the second term. Thus the Residue relation is guaranteed to be satisfied. Higher $1/J$ terms also create $\log z$, but they are associated with the mixing of higher twist double-stress-tensors. In this way, although we can say nothing about the mixed anomalous dimension associated with higher twist double-stress-tensors because of the lack of non-universal pieces in higher twist double-stress-tensors, we can insist that the Residue relation does hold true.

However, the above arguments can not be applied to the mixed OPE coefficients. It can be easily seen that different values of $J$ we integrate from brings up different non-logarithmic parts, reflecting that the mixed OPE coefficients are sensitive to lower spin region and thus the integration approximation should not be applied any more. We shall start with anomalous dimensions with finite spin obtained in section \ref{finitej} and sum them over to produce the correlator that tells us the mixed OPE coefficients, as we did in this section before.

We actually have another way to think about the Residue relation, similar to the arguments in \cite{Li:2019tpf}. Specifically, we can directly analyze the conformal blocks. We decompose the correlator into conformal blocks
\be
\mathcal{G}_{\mu^2}=\sum_J(c_{T^2} G_{2(d-2)+J,J}+\sum_n c_{n,J}\,G_{2\Delta_L+J+2n,J}+\cdots)\,,
\ee
where $c_{n,J}$ is the OPE coefficient associated with the double-trace operators $[\mathcal{O}_L\mathcal{O}_L]_{n,J}$ and dots represents higher twist double-stress-tensors and other possible operators. We consider the pole $\Delta^p_L$ of $c_{T^2}$ where $2\Delta^p_L-J+2n_p=2(d-2)$, then it is expected that $c_{n_p,J}$ contains the same pole $\Delta^p_L$ as discussed in \cite{Li:2019tpf,Fitzpatrick:2019zqz}. In general, we may expand $c_{T^2}$ and $c_{n_p,J}$ in terms of $\Delta_L-\Delta_L^p$
\bea
&& c_{T^2}=\fft{c_{T^2}^{-1}}{\Delta_L-\Delta_L^p}+c_{T^2}^0+c_{T^2}^1(\Delta_L-\Delta_L^p)+\cdots\,,
\cr &&
c_{n_p,J}=\fft{c_{n_p,J}^{-1}}{\Delta_L-\Delta_L^p}+c_{n_p,J}^0+c_{n_p,J}^1(\Delta_L-\Delta_L^p)+\cdots\,,
\eea
where $c_{T^2}^{-1}={\rm Res}_{\Delta_L=\Delta_L^p}\, c_{T^2}$. Then by taking $\Delta_L=\Delta_L^p$ we have
\be
\mathcal{G}_{\mu^2}=\sum_J\big((c_{n_p,J}^{0}+c_{T^2}^0)G_{2\Delta_L+J+2n_p,J}+\fft{c_{n_p,J}^{-1}+c_{T^2}^{-1}}{\Delta_L-
\Delta_L^p}G_{2\Delta_L+J+2n_p,J}+ c_{n_p,J}^{-1}G_{2\Delta_L+J+2n_p,J} \log(z \bar{z})\big)\,.
\ee
It is then straightforward to see $c_{n_p,J}^{-1}=c^{\rm mix}\gamma^{\rm mix}/2$. In addition, the correlator at $\Delta_L=\Delta_L^p$ should be well-defined, which requires us to cancel the divergence, leading to
\be
\gamma^{\rm mix}c^{\rm mix}=-2{\rm Res}_{\Delta_L=\Delta_L^p}\,c_{T^2}\,.
\ee
The mixed OPE coefficients are thus given by
\be
c^{\rm mix}=c_{n_p,J}^{0}+c_{T^2}^0\,.
\ee
We can actually clearly extract the contributions from the lowest-twist double-stress-tensor and double-trace operator separately! In Appendix \ref{mixOPE}, we extract the full mixed OPE for all pole examples in $d=6,8$. However, it is still not possible to recover the pure double-trace OPE coefficient before mixing, since the coefficients that vanish at $\Delta_L=\Delta_L^p$ are not known, which are, probably non-universal.
\section{Discussion on CFT$_2$}
\label{CFTtwo}

\cite{Li:2019zba} did not consider the case of $d=2$, but its framework can actually be applied to CFT$_2$. In the case $d=2$, one can immediately see that all the lowest-twist multi-stress-tensors have twist zero due to $d-2$ factor, as the result, there are no poles in $\Delta_L$. This is true, in $d=2$, there is no mixing happen to multi-stress-tensor. In fact, the enhanced conformal symmetry, i.e. the Virasoro symmetry, allows one to gather all lowest-twist multi-stress-tensors together to find the vacuum Virasoro conformal block \cite{Fitzpatrick:2014vua}, and moreover, there are no higher twist multi-stress-tensors.

In this section, we would like to discuss CFT$_2$ from the global Lorentzian inversion formula by following the framework of \cite{Li:2019zba}. By using the extracted finite spin $[\mathcal{O}_H\mathcal{O}_L]_{n,J}$, we shall then verify, up to double-stress-tensor,
the lowest-twist double-stress-tensor is the only thing we have, as consistent with Virasoro symmetry. We verify this fact by using the concept of twist conformal block \cite{Alday:2016njk}. These careful double-checks may allow us to gain some intuitions about how to connect the properties of OPE coefficients to Virasoro symmetry, which is beneficial for us to study the asymptotic Virasoro symmetry in high dimensions in the future.

\subsection{Finite spin double-twist operators}
The way to extract the OPE coefficients and anomalous dimensions with finite spin for double-twist operators $[\mathcal{O}_H\mathcal{O}_L]_{n,J}$ in $d=2$ is exactly the same as in $d=4$ in section \ref{finitej}, even much simpler, since the conformal block in CFT$_2$ is very simple
\be
G^{a,b}_{\Delta,J}(z,\bar{z})=k^{a,b}_{\Delta+J}(z)k_{\Delta-J}^{a,b}(\bar{z})+k^{a,b}_{\Delta-J}(z)k_{\Delta+J}^{a,b}(\bar{z})\,.
\ee
Thus the single-stress-tensor correlator is very simple
\be
\mathcal{G}_T=-\fft{\Delta_L}{2}\fft{(z\bar{z})^{\fft{\Delta_H+\Delta_L}{2}}}{((1-z)(1-\bar{z}))^{\Delta_L}}\big(1+\fft{1+z}{2(1-z)}\log z\big) + \big(z\leftrightarrow \bar{z}\big)\,.
\ee
We still decompose this correlator by looking at logarithmic part and non-logarithmic part separately for simplicity. For non-logarithmic part, we do not even really do the calculation, since it is nothing but the free correlator! We can immediately write the answer down
\be
\tilde{c}_{n,J}^{(1){\rm non-log}}=-\Delta_L\,,
\ee
without any dependence on $J$, since the $J$ dependent part is nothing but free OPE (\ref{freeOPEheav}). The logarithmic part (include both $\log z$ and $\log \bar{z}$) is not so trivial, but still not hard to work out. Follow the procedure in $d=4$ case, we find \footnote{As in $d=4$, the OPE result in previous version is also missing the appended term from another half of stress-tensor block.}
\be
\tilde{\gamma}^{(1)}_{n,J}=-\fft{1}{2}(\Delta_L+2n)\,,\qquad \tilde{c}^{(1)}_{n,J}=\fft{1-2\Delta_L}{2}-\fft{1}{2}J
\,(\psi_{n+J+1}-\psi_{n+J+\Delta_L})\,.
\ee
It follows that the anomalous dimension also does not have $J$ dependence at all! We will see in the next subsection that this is the essential origin of the uniqueness of lowest-twist double-stress-tensor.

\subsection{Uniqueness of double-stress-tensor}
In order to go to double-stress-tensor, we have to square the anomalous dimensions. We only keep the term that survives under dDisc
\be
\sum_{n=0,J=0}\tilde{c}^{\rm free}_{n,J} \fft{\tilde{\gamma}^{(1)2}_{n,J}}{8} (\log z)^2 z^{n+\fft{\Delta_H+\Delta_L}{2}}\bar{z}^{n+J+
\fft{\Delta_H+\Delta_L}{2}}
\,.\label{squreansadim2}
\ee
Since $\tilde{\gamma}^{(1)}_{n,J}$ contains no $J$, the above summation over $J$ is exactly same as in free field theory, which is defined as the twist conformal block \cite{Alday:2016njk}. In our case, we shall call it the heavy twist conformal block
\be
H_n(z,\bar{z}):= \sum_J \tilde{c}^{\rm free}_{n,J} \,G^{a,b}_{\Delta,J}(z,\bar{z})=\sum_J \tilde{c}^{\rm free}_{n,J}z^{n+\fft{\Delta_H+\Delta_L}{2}}\bar{z}^{n+J+
\fft{\Delta_H+\Delta_L}{2}} \,.
\ee
By doing the inversion job for free field theory, we can easily find the heavy twist conformal block
\be
H_{n}(z,\bar{z})=\fft{\Gamma(n+\Delta_L)}{\Gamma(n+1)\Gamma(\Delta_L)}\fft{z^{n+\fft{\Delta_H+\Delta_L}{2}}\bar{z}^{
\fft{\Delta_H+\Delta_L}{2}} }{(1-\bar{z})^{\Delta_L}}\,.
\ee
Thus the required correlator can be evaluated by
\be
\sum_{n=0,J=0}\fft{\tilde{\gamma}^{(1)2}_{n,J}}{8} (\log z)^2 H_n(z,\bar{z})\,,
\ee
and then go to the cross channel by $(z\rightarrow 1-\bar{z},\bar{z}\rightarrow 1-z)$. We end up with
\be
\mathcal{G}_{T^2}(z,\bar{z})\big|_{\log(1-\bar{z})^2}=\fft{\Delta_L\big(4(1-\bar{z})+\Delta_L(2-\bar{z})^2\big)}{32\bar{z}^2}\,.\label{corrdim2}
\ee
(\ref{corrdim2}) can be immediately verified to be true by comparing with the vacuum Virasoro conformal block \cite{Fitzpatrick:2014vua}
\be
\mathcal{V}_0=z^{\Delta_L} (1-z)^{-\fft{\Delta_L}{2}(1-\alpha)}\big(\fft{1-(1-z)^\alpha}{\alpha}\big)^{-\Delta_L}\,.\qquad \alpha=\sqrt{1-\mu}\,.\label{viraconf}
\ee
Expanding (\ref{viraconf}) in terms of $\mu$ and crossing it $(z\rightarrow 1-\bar{z},\bar{z}\rightarrow 1-z)$ and then extracting the term with $\mu^2$ and $\log(1-\bar{z})^2$, we can precisely obtain (\ref{corrdim2}). It is apparent from (\ref{corrdim2}) that it contains no $z$ dependence, implying it encodes and only encodes $n=0$ trajectory, i.e. the lowest-twist double-stress-tensor. We shall call this the uniqueness of double-stress-tensor in CFT$_2$. Using the Lorentzian inversion formula with putting (\ref{corrdim2}) in the dDisc, we can easily find the double-stress-tensor OPE given by
\be
c_{T^2}=\fft{2^{-1-2J}\sqrt{\pi}\Delta_L \big(4+(J^2-J-2)\Delta_L\big)\Gamma(J+1)}{(J-3)J(J+2)\Gamma(J-\fft{1}{2})}\,,
\ee
exactly agrees with one found in \cite{Kulaxizi:2019tkd}.

In CFT$_2$, it is understood that this uniqueness and universality of double-stress-tensor is guaranteed by the Virasoro symmetry, but from the global inversion perspective, it appears that the $J$ independent anomalous dimensions should be responsible for the uniqueness and universality. In fact, the origin of this $J$-independence is the $\bar{z}$-independence of the (a half of) crossed single-stress-tensor conformal block. Looking at the single-stress-tensor conformal block in $d=2$, we can easily find that $z$-dependence and $\bar{z}$-dependence is completely factorized. This implies that the underlying symmetry can be factorized to pure holomorphic and pure anti-holomorphic part, exactly as Virasoro symmetry indicates. The uniqueness of double-stress-tensor can be directly understood from this holomorphic factorization: higher twists $n>0$ are coming with the derivative $\partial_{z}\partial_{\bar{z}}$ that mixes the holomorphic sector and anti-holomorphic sector, thus they are enforced to be vanishing, and the universality comes with this uniqueness.

So the question is then what can we learn about higher dimensional version of Virasoro-like structure analogous to $d=2$? We should start from the properties of double-stress-tensor OPE in high dimension. There is no uniqueness at all, which simply implies that CFT in high dimension does not have the nice holomorphic factorization. Nevertheless, the lowest-twist double-stress-tensor shows the universality (ignoring the case that $\Delta_L$ takes values at poles), indicating that the lowest-twist double-stress-tensor might be controlled by the symmetry similar to Virasoro symmetry. We can imagine that some operators $(L_m, \bar{L}_m)$ can be constructed from stress-tensor operator by following the construction in $d=2$, but in general
\be
[L_m,\bar{L}_n]\neq0\,.
\ee
However, it is plausible that near the lightcone limit $L_m$ and $\bar{L}_n$ is asymptotically commutable
\be
L_m=L^{(0)}_m+\cdots\,,\qquad \bar{L}_m=\bar{L}^{(0)}_m+\cdots\,,\qquad [L^{(0)}_m,\bar{L}^{(0)}_n]=0\,,
\ee
where dots represent terms that are suppressed in the lightcone limit. Then the commutable subsectors $L^{(0)}_m$ form Virasoro-like algebra that controls lowest-twist multi-stress-tensors. This asymptotic Virasoro structure was constructed recently in \cite{Huang:2019fog,Huang:2020ycs}.

\section{Conclusion}
\label{conclu}
In this paper, we continued studying heavy-light four-point function up to double-stress-tensor, providing the supplements to previous work \cite{Li:2019zba}.

We analyzed the leading and next-to-leading Regge behavior (in the large impact regime) of lowest-twist double-stress-tensor in general even dimensions by noting the Regge poles in the lowest-twist double-stress-tensor OPE. Then we used the Lorentzian inversion formula to extend the universality of double-twist operators $[\mathcal{O}_H\mathcal{O}_L]_{n,J}$ at the linear order $\mathcal{O}(\mu)$ from large spin to finite spin. More specifically, we extracted the OPE coefficients and anomalous dimension of double-twist operators  with finite spin by using the Lorentzian inversion formula where the large spin expansion of our results agree with those found in \cite{Karlsson:2020ghx}. We also took advantage of geodesic Witten diagram to find $\bar{z}\rightarrow 1$ expansion for the logarithmic part of single-stress-tensor block in general dimension, which allowed us to bootstrap the anomalous dimension of double-twist operators with finite spin in any dimension. The fact that these anomalous dimensions with finite spin are consistent with holographic computations were observed.

By using these anomalous dimension with finite spin, we addressed $\Delta_L$ poles. Specifying $\Delta_L$ to specific poles, we found that we can summed over spin and twist to obtain $\mathcal{O}(\mu^2)$ correlators without divergence in $\Delta_L$, from which it is obvious to see the mixing between double-trace operators $[\mathcal{O}_L\mathcal{O}_L]_{n,J}$ with certain twist $n_0$ and the lowest-twist double-stress-tensor. The mixed anomalous dimension arises in the mixing trajectory and we verified that the product of the mixed anomalous dimension and the mixed OPE coefficient obeys the Residue relation quoted in \cite{Li:2019tpf}. The universality of this Residue relation was also understood. Besides, the double-trace OPE coefficients with $n\leq n_0$ and the mixed OPE coefficients associated with all poles in $d=4,6$ were computed. Furthermore, we considered CFT$_2$ within the framework of \cite{Li:2019zba} and this paper. We extracted the OPE coefficients and anomalous dimensions of double-twist operators $[\mathcal{O}_L\mathcal{O}_L]_{n,J}$ with finite spin in $d=2$, and in particular the anomalous dimensions are actually independent of $J$. From $J$-independent anomalous dimension, we found we were led to the uniqueness of double-stress-tensor. In other words, the lowest-twist double-stress-tensor is only thing exists in the double-stress-tensor sector of heavy-light four-point function in $d=2$, which is consistent with Virasoro symmetry \cite{Fitzpatrick:2014vua}.

There are other tantalizing questions to investigate in the future. It is especially interesting to think about how to study heavy-light four-point function from AdS/CFT in a more formal way, verifying the validity of the framework proposed in \cite{Fitzpatrick:2019zqz}. In other words, how do we take the heavy-limit of Witten diagram in general higher derivative gravities? This question could also shed light on the question of what is the scattering amplitudes corresponding to the flat-space limit of heavy-light four-point function.

\section*{Acknowledgement}
We are grateful to Nikhil Anand, Simon Caron-Huot, Wenliang Li and Hong Lu for helpful discussions. We are grateful to Alba Grassi, Cristoforo Iossa, Daniel Panea Lichting, Matthew Dodelson and Sasha Zhiboedov for pointing out the error of $d=4$ finite spin OPE result in previous version of this paper. This work is supported in part by the NSFC (National Natural Science Foundation of China) Grant No. 11935009 and No.~11875200.

\appendix

\section{Lorentzian inversion formula}
\label{inversion}
In this Appendix, we briefly review the Lorentzian inversion formula \cite{Caron-Huot:2017vep,Simmons-Duffin:2017nub,Kravchuk:2018htv}
\be
c(\Delta,J)=\fft{1+(-1)^J}{4}\kappa^{a,b}_{\Delta+J}\int dzd\bar{z}\,\mu^{a,b}(z,\bar{z})G^{a,b}_{J+d-1,\Delta-d+1}(z,\bar{z}){\rm dDisc}[\mathcal{G}(z,\bar{z})]\,,\label{Loreninv}
\ee
where
\bea
&& \mu^{a,b}(z,\bar{z})=\Big|\fft{z-\bar{z}}{z\bar{z}}\Big|^{d-2}\fft{\big((1-z)(1-\bar{z})\big)^{a+b}}{(z\bar{z})^2}\,,
\cr && \kappa^{a,b}_{\beta}=\fft{\Gamma(\fft{\beta}{2}-a)\Gamma(\fft{\beta}{2}+a)\Gamma(\fft{\beta}{2}-b)\Gamma(\fft{\beta}{2}+b)}{2\pi^2 \Gamma(\beta-1)
\Gamma(\beta)}\,.
\eea
and ${\rm dDisc}$ is the the double-discontinuity given by the double commutator $\langle[\mathcal{O}_1\mathcal{O}_4][\mathcal{O}_2\mathcal{O}_3]\rangle$ that can be evaluated by
\be
{\rm dDisc}[\mathcal{G}(z,\bar{z})]=\cos(\pi(a+b))\mathcal{G}(z,\bar{z})-\fft{e^{-i(a+b)}}{2}\mathcal{G}^{\circlearrowleft}(z,\bar{z})
-\fft{e^{i(a+b)}}{2}\mathcal{G}^{\circlearrowright}(z,\bar{z})\,,
\ee
where $\mathcal{G}^{\circlearrowleft}$ and $\mathcal{G}^{\circlearrowright}$ are two different analytic continuations for $\bar{z}$ around $1$. We can set $(-1)^{J}=1$ in this paper since the exchanged operators are expected to have even spin. The physical spectrum and the corresponding OPE coefficients are encoded in the OPE data $c(\Delta,J)$ \cite{Caron-Huot:2017vep}
\be
c(\Delta,J)=-\fft{c_{\Delta_{\rm phys},J}}{\Delta-\Delta_{\rm phys}}\,,\label{cOPEdata}
\ee
where $\Delta_{\rm phys}$ is the physical spectrum including the anomalous dimension
\be
\Delta_{\rm phys}=\Delta_{\rm bare}+\gamma_{\rm phys}\,,
\ee
where $\Delta_{\rm bare}$ is the bare physical spectrum at the degenerate point or in free field theory.

In the end, there is an integral formula that is used throughout this paper  \cite{Caron-Huot:2017vep}
\bea
&& I^{a,b}_{\hat{\tau}}(\beta)=\int_0^1 \fft{d\bar{z}}{\bar{z}^2}(1-\bar{z})^{a+b}\kappa^{a,b}_\beta k^{a,b}_\beta(\bar{z})\,{\rm dDisc}[
\big(\fft{1-\bar{z}}{\bar{z}}\big)^{\fft{\hat{\tau}}{2}-b}(\bar{z})^{-b}]
\cr &&
\cr && =\fft{\Gamma(\fft{\beta}{2}-a)\Gamma(\fft{\beta}{2}+b)\Gamma(\fft{\beta}{2}-\fft{\hat{\tau}}{2})}
{\Gamma(-\fft{\hat{\tau}}{2}-a)\Gamma(-\fft{\hat{\tau}}{2}+b)
\Gamma(\beta-1)\Gamma(\fft{\beta}{2}+\fft{\hat{\tau}}{2}+1)}\,.\label{intresult}
\eea

\section{Series expansion of conformal blocks}
\label{seriesexp}
In this Appendix, we review the series expansion of conformal block with using the Casimir operators \cite{Hogervorst:2013sma}.

We consider the series expansion of conformal blocks in $u=z\bar{z}\rightarrow 0$ \cite{Hogervorst:2013sma}
\be
G^{a,b}_{\Delta,J}=\sum_{n=0}G^{(n)}_{\Delta,J}=\sum_{m=-n}^n \mathcal{A}^{a,b}_{nm}\mathcal{P}_{\Delta+n,J+m}(z,\bar{z})\,,\label{seexpand}
\ee
where $\mathcal{P}_{\Delta,J}(z,\bar{z})$ can be simply expressed as the Gegenbauer polynomial
\bea
&& \mathcal{P}_{\Delta,J}(z,\bar{z})=\fft{2^{3-d-J}\sqrt{\pi}\Gamma(J+d-2)}{\Gamma(\fft{d-1}{2})\Gamma(\fft{d}{2}+J-1)}
(z\bar{z})^{\fft{\Delta}{2}}\tilde{C}_J(\fft{z+\bar{z}}{2\sqrt{z\bar{z}}})\,,
\cr &&
\cr && \tilde{C}_J(x)=\fft{\Gamma(d+J-2)}{\Gamma(d-2)\Gamma(J+1)}C_J^{\fft{d}{2}-1}(x)=\, _2F_1(-J,d+J-2,\frac{d-1}{2},\frac{1-x}{2})\,.
\eea
Then the task is to find out the coefficients $\mathcal{A}^{a,b}_{nm}$. The trick is to use the (quadratic) Casimir operator, since the conformal block is the eigenfunction of the Casimir operator, i.e.
\bea
&& \mathcal{C}_2 \,G^{a,b}_{\Delta,J}(z,\bar{z})
=C_{\Delta,J}\,G^{a,b}_{\Delta,J}(z,\bar{z})\,,
\cr && C_{\Delta,J}=\Delta(\Delta-d)+J(J+d-2)\,,\label{Casi2}
\eea
where
\bea
&& \mathcal{C}_2=\mathcal{D}_z+\mathcal{D}_{\bar{z}}+2(d-2)\fft{z\bar{z}}{z-\bar{z}}
((1-z)\partial_z-(1-\bar{z})\partial_{\bar{z}})\,,
\cr && \mathcal{D}_z=2(z^2(1-z)\partial_z^2-(1+a+b)z^2\partial_z-abz)\,.\label{Casi2full}
\eea
In fact, it turns out that the Casimir operator $\mathcal{C}_2$ can be separated into two parts $\mathcal{C}_2^1$ and $\mathcal{C}_2^2$ where $\mathcal{C}_2^1$ is homogeneous in $u$ and $\mathcal{C}_2^2$ shifts $u$ by one unit \cite{Hogervorst:2013sma}. Specifically, it was found \cite{Hogervorst:2013sma}
\bea
\mathcal{C}_2^1 \mathcal{P}^{a,b}_{\Delta,J}=C_{\Delta,J} \mathcal{P}^{a,b}_{\Delta,J}\,,\qquad \mathcal{C}_2^2 \mathcal{P}^{a,b}_{\Delta,J}=\gamma^{-,a,b}_{\Delta,J} \mathcal{P}^{a,b}_{\Delta+1,J-1}+\gamma^{+,a,b}_{\Delta,J} \mathcal{P}^{a,b}_{\Delta+1,J+1}\,,
\eea
where
\bea
&& \gamma^{+,a,b}_{\Delta,J}=(\Delta+J+2a)(\Delta+J+2b)\,,
\cr && \gamma^{-,a,b}_{\Delta,J}=\frac{J (d+J-3) (-2 a+d-\Delta +J-2) (-2 b+d-\Delta +J-2)}{(d+2 J-4) (d+2 J-2)}\,.
\eea
Then one can have the recursion equation for determining $\mathcal{A}^{a,b}_{nm}$
\be
(C_{\Delta+n,J+m}-C_{\Delta,J})\mathcal{P}^{a,b}_{\Delta,J}=\gamma^{+,a,b}_{\Delta+n-1,J+m-1}\mathcal{P}^{a,b}_{\Delta+n-1,J+m-1}
+\gamma^{-,a,b}_{\Delta+n-1,J+m+1}\mathcal{P}^{a,b}_{\Delta+n-1,J+m+1}\,.\label{recurseCas}
\ee

\section{Stress-tensor block from geodesic Witten diagram}
\label{geowitten}
In this Appendix, we would briefly review spin-$2$ geodesic Witten diagram that serves as the integral representation of the stress-tensor conformal block \cite{Hijano:2015zsa} and then we would describe the trick how we derive the logarithmic part in $\bar{y}$ expansion of stress-tensor conformal block (\ref{stresszb11}), (\ref{stresszb12}).

The geodesic Witten diagram is defined holographically in AdS$_{d+1}$
\bea
&& \mathcal{W}_{\Delta,J}(x_i) =\int_{\gamma_{12},\gamma_{34}}G_{b\partial}(y(\lambda),x_1)G_{b\partial}(y(\lambda),x_2)G_{bb}(y(\lambda),y(\lambda');
\Delta,J)G_{b\partial}(y(\lambda'),x_3)G_{b\partial}(y(\lambda'),x_4)\,,
\cr &&\label{geoWitt}
\eea
where $G_{b\partial}$ and $G_{bb}$ are bulk-to-boundary and bulk-to-bulk propagators respectively. Roughly speaking, the geodesic Witten diagram is just Witten diagram but it integrates along the geodesics that connect the boundary points $x_{1,3}$ and $x_{2,4}$. It is proved that the geodesic Witten diagram satisfies the Casimir equation and thus should be viewed as the conformal block with exchanging $(\Delta,J)$. In our case $(\Delta=d,J=2)$ where (denote $y(\lambda')=y'$)
\be
G_{bb}\sim (G_{bb})_{\mu\nu\rho\sigma}\fft{\partial y^\mu}{\partial\lambda}\fft{\partial y^\nu}{\partial\lambda}
\fft{\partial y'^\rho}{\partial\lambda'}\fft{\partial y'^\sigma}{\partial\lambda'}\,.
\ee

We are mainly interested in the case with $a=b=0$, and one may set this from now on such that the calculation would be more smooth, however, this is not permitted, since setting $a=b=0$ in (\ref{geoWitt}) would cause the divergence of the integration. The right way is to work out (\ref{geoWitt}) at first and then set $a=b=0$, which is rather complicated. Let us not be so ambitious, we are only interested in the logarithmic part of the conformal block which is closely related to the fake divergence, thus the trick is to regulate the divergence in $a$ or $b$ and its coefficients shall be what we are looking for. For simplicity, we may set $b=0$ and work on (\ref{geoWitt}) to find the regulated divergence $1/a$. Within the conformal frame $(z,\bar{z})$, we can parametrize the geodesics as \cite{Hijano:2015zsa}
\bea
&& ds^2=\fft{1}{u^2}(du^2+dzd\bar{z}+\cdots)\,,\qquad u=\fft{\sqrt{z \bar{z}}}{2\cosh\lambda}\,,
\cr &&
\cr && z(\lambda)=\fft{2-z}{2}-\fft{z}{2}\tanh\lambda\,,\qquad  \bar{z}(\lambda)=\fft{2-\bar{z}}{2}-\fft{\bar{z}}{2}\tanh\lambda\,,
\eea
then we find
\be
\mathcal{W}_{d,2}=\int_0^1 d\sigma \ft{\Gamma(d+2)}{\Gamma(1-a+\fft{d}{2})\Gamma(1+a+\fft{d}{2})}\ft{(z\bar{z})^{\fft{d-2}{2}}((1-z\sigma)(1
-\bar{z}\sigma))^{\fft{d-2}{2}}\sigma^{\fft{d}{2}+a}(1-\sigma)^{\fft{d}{2}-a}}{d(1-z(1-\bar{z})\sigma-\bar{z}\sigma)^2}
\mathcal{F}(z,\bar{z},\sigma)\,,\label{Wd2}
\ee
where
\bea
&& \mathcal{F}(z,\bar{z},\sigma)=d (\sigma ^2 z^4 (\bar{z}-1)^2 (2 \sigma  \bar{z} (\sigma  \bar{z}-1)+1)-2 \sigma  z^3 (\bar{z}-1) (\sigma  \bar{z} (\sigma  \bar{z}-1) (2 \sigma  \bar{z}+\bar{z}-3)+\bar{z}-1)
\cr && +z^2 (\bar{z} (-6 \sigma +\bar{z} (\sigma  (14 \sigma +\bar{z} (\sigma  (-2 (5 \sigma +4)+2 \sigma  (\sigma +2) \bar{z}+\bar{z})-2)+4)+2)-2)+1)
\cr &&-2 z \bar{z} (\sigma  \bar{z}-1)^2 (\sigma  \bar{z}+\bar{z}-1)+\bar{z}^2 (\sigma  \bar{z}-1)^2)
 -2 z \bar{z} (\sigma  z-1) (\sigma  \bar{z}-1) (\sigma ^2 (z (-\bar{z})+z+\bar{z})^2
 \cr &&+2 \sigma  z (\bar{z}-1)+z (\bar{z}-1)-2 \sigma  \bar{z}-\bar{z}+2)\,,
\eea
with $\sigma=e^{2\lambda'}/(1+e^{2\lambda'})$.

Then our trick of the guessing job to find (\ref{stresszb11}) is as follows
\begin{itemize}
\item Expand (\ref{Wd2}) as the series of $\bar{y}$ at first, integrate it over $\sigma$ and then send $a\rightarrow 0$ to keep terms with $1/a$. This remaining term with $1/a$ shall be the logarithmic part, indicating the replacement $1/a\rightarrow \log\bar{y}$. Expand the resulting expressions as the series of $y$ for each $\bar{y}^k$, it is made possible to find patterns for the coefficients associated with $y^m$ and in the end the pattern for coefficients $C_{km}$ associated with $y^m \bar{y}^k$ can be found. With $C_{km}$ in hand, (\ref{stresszb11}) can be deduced by
    \be
    \sum_{k}\sum_{m=0}^k C_{km}y^m \bar{y}^k \log\bar{y}\,.
    \ee

\end{itemize}

\section{$[\mathcal{O}_H\mathcal{O}_L]_{n,J}$ from holography}
\label{holography}
In this Appendix, we extract the double-twist anomalous dimensions with finite spin by using the holographic Hamiltonian perturbation theory \cite{Kaviraj:2015xsa,Fitzpatrick:2014vua}, and it is observed that the holographic results precisely agree with the results in the main text from Lorentzian inversion formula.

From the holographic perspective, the conformal dimension of $[\mathcal{O}_1\mathcal{O}_2]_{n,J}$ is interpreted as the energy of the bound state of two particles $1$ and $2$ circling around each other. In the heavy-limit we consider, the picture is that the light particle is circling around the black hole, and the anomalous dimension is the energy-shift of that particle caused by the interaction between itself and the black hole. We consider the general higher derivative gravity minimally coupled to massive scalars in bulk, giving the Lagrangian as
\bea
L=R-2\Lambda+H(R_{\mu\nu\rho\sigma})-\frac{1}{2}(\partial_\mu \phi \partial^\mu \phi+ \Delta_L(\Delta_L-d)\phi^2)\,,
\eea
where $H(R_{\mu\nu\rho\sigma})$ is the generic higher derivative terms. We consider the spherical symmetric black hole constructed from the gravity sector and the scalar fields are considered as the field that creates particles moving around such a black hole. Typically, we should consider the massless higher derivative gravity where the massive modes are decoupled \cite{Li:2018drw} to guarantee there is no extra operators as input in addition to single-stress-tensor \cite{Li:2019tpf}. In this way, by dimensional analysis, the general metric ansatz is
\be
ds^2=f dt^2+h^{-1}dr^2+r^2 d\Omega_{d-1}^2\,,
\ee
where
\bea
&&f=1+r^2-\frac{\mu}{r^{d-2}}+\sum_{k=2}^\infty \frac{\mu^k f_k}{r^{k d-2}}\crcr
&&h=1+r^2-\frac{\mu}{r^{d-2}}+\sum_{k=2}^\infty \frac{\mu^k h_k}{r^{k d-2}}\,,
\label{met}
\eea
where $f_k$ and $h_k$ depend on the detail of higher derivative terms. To compute the anomalous dimension, we should just do quantum mechanics on this black hole background: write down the Hamiltonian $H=H_0+H_{\rm int}$ where $H_0$ is the free Hamiltonian and $H_{\rm int}$ represents the interaction, and then compute $\langle n,J| H_{\rm int}|n,J\rangle$. Up to the order $\mathcal{O}(\mu)$, we find
\bea
H_{\rm int}=\mu \int d\Omega_{d-1}dr \bigg( -\frac{r \dot{\phi}^2 }{2(1+r^2)^2}-\frac{r}{2}(\partial_r\phi)^2\bigg)\,,
\eea
where all theory dependent terms $f_k$ make no contributions, implies that the anomalous dimensions of double-twist operators are exactly universal at the order $\mathcal{O}(\mu)$. The quantization can be standardly performed
\be
\phi=\sum_{n,J,j}(a^\dagger_{n,J,j}\psi_{n,J,j}+a_{n,J,j}\psi^*_{n,J,j})\,,
\label{phiquantize}
\ee
where $a^\dagger$ and $a$ are creation operator and annihilation operator respectively and the mode function $\psi_{n,J,j}$ is given by \cite{Fitzpatrick:2010zm}
\be
\psi_{n,J,j}(t,r,\Omega)=\frac{1}{N_{\Delta_L,n,J}} e^{- i E_{n,J} t} Y_{J,j}(\Omega) \frac{r^J}{(1+r^2)^{\frac{\Delta_L+J}{2}}}\,_2F_1\Big(-n,\Delta_L+J+n,J+\frac{d}{2},\frac{r^2}{r^2+1}\Big)\,,\label{modefunc}
\ee
where
\bea
&&E_{n,J}=\Delta_L + 2n +J \,,\crcr
&& N_{\Delta,n,J}=(-1)^n \bigg (\frac{n! \Gamma(J+\frac{d}{2}) \Gamma(\Delta+n-\frac{d-2}{2})}{\Gamma(n+J+\frac{d}{2})\Gamma(\Delta+n+J)}\bigg) ^{\fft{1}{2}}\,.
\eea
It is then straightforward to compute the anomalous dimension by
\bea
&&\mu \tilde{\gamma}^{(1)}_{n,J}=\langle n,J,j | H_{\rm int} | n,J,j\rangle\,, \crcr
&&=\mu \int dr d\Omega \bigg( -\frac{r \partial_t \psi_{n,J,j} \partial_t \psi_{n,J,j}^*}{(1+r^2)^2}- r \partial_r\psi_{n,J,j}\partial_r\psi_{n,J,j}^*\bigg)\,.\label{tocalano}
\eea
In the literature, it is normal to just keep $\partial_t \psi_{n,J,j} \partial_t \psi_{n,J,j}^*$ and drop $\partial_r\psi_{n,J,j}\partial_r\psi_{n,J,j}^*$, because $\partial_r\psi_{n,J,j}\partial_r\psi_{n,J,j}^*$ only contributes $1/J$ correction \cite{Fitzpatrick:2014vua,Kaviraj:2015xsa,Kulaxizi:2018dxo}. Substituting (\ref{modefunc}) into (\ref{tocalano}), we can work out the integral $n$ by $n$ and find the pattern, and the results precisely agree with the anomalous dimensions obtained in the main text for $d=4,6,8,10$ (\ref{dim4gaope}), (\ref{d6}), (\ref{d8}) and (\ref{d10}).
\section{Double-trace and mixed OPE coeffcieints}
\label{mixOPE}
The expressions of mixed OPE coefficients and the OPE coefficients associated with those double-trace operators (without mixing with double-stress-tensor) are extremely complicated, thus we put what we find for $d=6$ and $d=8$ as examples in this Appendix. The complexity of these OPE coefficients are resulted from complicated double-twist anomalous dimensions in section \ref{finitej}: the resulting correlator with summing $n$ and $J$ over would contain transcendental functions. In particular, there will be Lerch transcendents $\Phi\big((1-z)^2,2,a/2\big)$ where $a$ is odd integer. These functions can actually be decomposed into some PolyLog functions ${\rm Li}_2$, for examples
\bea
&& \Phi(x,2,\ft{1}{2})=-\frac{2 \left(\text{Li}_2\left(-\sqrt{x}\right)-\text{Li}_2\left(\sqrt{x}\right)\right)}{\sqrt{x}}\,,
\cr &&
\cr && \Phi(x,2,\ft{3}{2})= -\frac{2 \left(\text{Li}_2\left(-\sqrt{x}\right)-\text{Li}_2\left(\sqrt{x}\right)+2 \sqrt{x}\right)}{x^{3/2}}\,,
\cr &&
\cr && \Phi(x,2,\ft{5}{2})=-\frac{2 \left(9 \text{Li}_2\left(-\sqrt{x}\right)-9 \text{Li}_2\left(\sqrt{x}\right)+2 (x+9) \sqrt{x}\right)}{9 x^{5/2}}\,.
\eea
Although the expressions of correlators are quite complicated, we still can expand the correlators in terms of $z$ up to the order where lowest-twist double-stress-tensor appears and then do the inversion integral to obtain $c(\Delta,J)$. It is convenient to expand $c(\Delta,J)$ in terms of $n=1/2(\beta-2J-\tau_0)\rightarrow 0$ where $\tau_0$ is the physical twist and $\beta=\Delta+J$, to write (\ref{cOPEdata}) as
\be
c(\Delta,J)=\fft{S_{-1}(\beta)}{n}+\fft{S_{-2}(\beta)}{n^2}+\cdots\,,
\ee
where the coefficients $S_{-1}(\beta)$ and $S_{-2}(\beta)$ are related to OPE coefficients and anomalous dimensions by
\bea
c_{\tau_0,J}=-2(S_{-1}+2\partial_\beta S_{-2})\big|_{\beta=\tau_0+2J}\,,\qquad c_{\tau_0,J}\gamma_{\tau_0,J}=-4S_{-2}\,.\label{relacandano}
\eea
Below we would just present $S_{-1}$ and $S_{-2}$ for all examples we consider. Particularly, it would be useful to denote
\be
\mathcal{I}_a=\fft{2^{4-\beta}\Gamma(a+1)^2 \Gamma(\fft{\beta}{2}-a-1)\Gamma(\fft{\beta}{2})}{\pi^{\fft{3}{2}}\Gamma(\fft{\beta-1}{2})\Gamma(\fft{\beta}{2}+a+1)}\,.
\ee

\subsection{$d=6$}
In $d=6$, there are two poles $\Delta_L=4,3$ \cite{Li:2019zba,Karlsson:2019dbd}.

\subsubsection{$\Delta_L=4$}

For $\Delta_L=4$, the lowest-twist double-stress-tensor mixes with leading-twist double-trace operator $[\mathcal{O}_L\mathcal{O}_L]_{0,J}$. The mixed OPE data is recorded as
\bea
S_{-2}=\frac{3}{2} \pi ^2 \left(\mathcal{I}_0+96 \mathcal{I}_1+1296 \mathcal{I}_2+5900 \mathcal{I}_3+11700 \mathcal{I}_4+10500 \mathcal{I}_5+3500 \mathcal{I}_6\right)\,,\label{Sn2dim6}
\eea
and
\bea
&& S_{-1}=6 \pi ^2 (21 \mathcal{I}_1+2215 \mathcal{I}_2+46246 \mathcal{I}_3+383283 \mathcal{I}_4+1626920 \mathcal{I}_5+3960290 \mathcal{I}_6+5772550 \mathcal{I}_7
\cr && +4989950 \mathcal{I}_8+2362500 \mathcal{I}_9+472500 \mathcal{I}_{10})-\frac{3}{2} \pi ^2 \log 2 \,(3 \mathcal{I}_0+744 \mathcal{I}_1+28024 \mathcal{I}_2+381740 \mathcal{I}_3
\cr && +2557220 \mathcal{I}_4+9695140 \mathcal{I}_5+22217580 \mathcal{I}_6+31393600 \mathcal{I}_7+26748400 \mathcal{I}_8+12600000 \mathcal{I}_9
\cr && +2520000 \mathcal{I}_{10})\,.
\eea
It is obvious that (\ref{Sn2dim6}) gives us (\ref{canodim6}) that satisfies the Residue relation.

\subsubsection{$\Delta_L=3$}
For $\Delta_L=3$, the lowest-twist double-stress-tensor mixes with sub-leading-twist double-trace, we then would like to extract both the leading-twist (non-mixing) double-trace OPE and the mixed OPE.
\begin{itemize}
\item[1.] Leading-twist
\bea
&& S_{-1}= \frac{9}{16} \pi ^2 \big((31 \mathcal{I}_1+2490 \mathcal{I}_2+39276 \mathcal{I}_3+246200 \mathcal{I}_4+783384 \mathcal{I}_5+1393200 \mathcal{I}_6+1407200 \mathcal{I}_7
\cr && +756000 \mathcal{I}_8+168000 \mathcal{I}_9)-\log2\,( \mathcal{I}_0+226 \mathcal{I}_1+6888 \mathcal{I}_2+74112 \mathcal{I}_3+387080 \mathcal{I}_4+1123632 \mathcal{I}_5
\cr && +1909600 \mathcal{I}_6+1889600 \mathcal{I}_7+1008000 \mathcal{I}_8+224000 \mathcal{I}_9)\big)\,.
\eea
There is no $S_{-2}$, reflects that the anomalous dimension does not exist and thus there is no mixing with stress-tensor at all.

\item[2.] Mixed
\bea
&& S_{-2}=-\frac{9}{32}\pi ^2 \left(\mathcal{I}_0+168 \mathcal{I}_1+2688 \mathcal{I}_2+13040 \mathcal{I}_3+26520 \mathcal{I}_4+24000 \mathcal{I}_5+8000 \mathcal{I}_6\right)\,.
\cr &&
\cr && S_{-1}=\frac{9}{32} \pi ^2(\mathcal{I}_0+288 \mathcal{I}_1+7016 \mathcal{I}_2+1316 \mathcal{I}_3-581016 \mathcal{I}_4-3953936 \mathcal{I}_5-12017536 \mathcal{I}_6
\cr && -20008800 \mathcal{I}_7-18929600 \mathcal{I}_8-9576000 \mathcal{I}_9-2016000 \mathcal{I}_{10})-\frac{27}{32} \pi ^2 (\mathcal{I}_0+98 \mathcal{I}_1-296 \mathcal{I}_2
\cr && -40576 \mathcal{I}_3-443304 \mathcal{I}_4-2160496 \mathcal{I}_5-5800576 \mathcal{I}_6-9156800 \mathcal{I}_7-8473600 \mathcal{I}_8-4256000 \mathcal{I}_9
\cr && -896000 \mathcal{I}_{10})\,.
\eea
It can be readily verified that the Residue relation holds true.
\end{itemize}

\subsection{$d=8$}
In $d=8$, we have poles located at $\Delta_L=6, 5, 4$ \cite{Li:2019zba}.

\subsubsection{$\Delta_L=6$}
For $\Delta_L=6$, the lowest-twist double-stress-tensor mixes with leading-twist double-trace, and we find
\bea
&& S_{-2}=\frac{45}{8} \pi ^2 (3 \mathcal{I}_0+520 \mathcal{I}_1+13390 \mathcal{I}_2+119820 \mathcal{I}_3+500910 \mathcal{I}_4+1105440 \mathcal{I}_5+1328880 \mathcal{I}_6+823200 \mathcal{I}_7
\cr && +205800 \mathcal{I}_8)\,,
\cr &&
\cr && S_{-1}=\frac{15}{4} \pi ^2(9 \mathcal{I}_0+14121 \mathcal{I}_1+2084124 \mathcal{I}_2+91807278 \mathcal{I}_3+1856575540 \mathcal{I}_4+21210703202 \mathcal{I}_5
\cr && +153524501602 \mathcal{I}_6+753615785692 \mathcal{I}_7+2614971373672 \mathcal{I}_8+6571097962296 \mathcal{I}_9+12095797182672 \mathcal{I}_{10}
\cr && +16315061347200 \mathcal{I}_{11}+15939178794000 \mathcal{I}_{12}+10974472821312 \mathcal{I}_{13}+5048455828416 \mathcal{I}_{14}
\cr && +1392269598720 \mathcal{I}_{15}+174033699840 \mathcal{I}_{16})-\frac{45}{8} \pi ^2 \log2\,(39 \mathcal{I}_0+26080 \mathcal{I}_1+2709366 \mathcal{I}_2+101741580 \mathcal{I}_3
\cr && +1907236670 \mathcal{I}_4+21018198144 \mathcal{I}_5+149649169616 \mathcal{I}_6+729638532000 \mathcal{I}_7+2526439025544 \mathcal{I}_8
\cr && +6348500905152 \mathcal{I}_9+11695090680000 \mathcal{I}_{10}+15789391123968 \mathcal{I}_{11}+15438335996160 \mathcal{I}_{12}
\cr && +10636095934464 \mathcal{I}_{13}+4894641303552 \mathcal{I}_{14}+1350079610880 \mathcal{I}_{15}+168759951360 \mathcal{I}_{16})\,.
\eea

\subsubsection{$\Delta_L=5$}
For $\Delta_L=5$, what the lowest double-stress-tensor mixes with is sub-leading twist double-trace.
\begin{itemize}
\item[1.] Leading-twist
\bea
&& S_{-1}=\frac{25}{8} \pi ^2 (3 \mathcal{I}_0+3228 \mathcal{I}_1+378804 \mathcal{I}_2+13872662 \mathcal{I}_3+237168194 \mathcal{I}_4+2305126614 \mathcal{I}_5
\cr && +14206144050 \mathcal{I}_6+59189235400 \mathcal{I}_7+173118036924 \mathcal{I}_8+362574882068 \mathcal{I}_9+546923130012 \mathcal{I}_{10}
\cr && +589539816936 \mathcal{I}_{11}+443012361792 \mathcal{I}_{12}+220435235328 \mathcal{I}_{13}+65262637440 \mathcal{I}_{14}
\cr && +8701684992 \mathcal{I}_{15})-\frac{75}{2} \pi ^2 \log2\,(\mathcal{I}_0+622 \mathcal{I}_1+56850 \mathcal{I}_2+1849522 \mathcal{I}_3+29874024 \mathcal{I}_4
\cr && +282767886 \mathcal{I}_5+1722915726 \mathcal{I}_6+7149558660 \mathcal{I}_7+20897483544 \mathcal{I}_8+43797760608 \mathcal{I}_9
\cr && +66136501200 \mathcal{I}_{10}+71359870176 \mathcal{I}_{11}+53663081472 \mathcal{I}_{12}+26713978368 \mathcal{I}_{13}+7910622720 \mathcal{I}_{14}
\cr && +1054749696 \mathcal{I}_{15})\,.
\eea

\item[2.] Mixed
\bea
&& S_{-2}=-\frac{75}{8} \pi ^2 (\mathcal{I}_0+240 \mathcal{I}_1+7080 \mathcal{I}_2+67580 \mathcal{I}_3+292020 \mathcal{I}_4+655620 \mathcal{I}_5+794780 \mathcal{I}_6+493920 \mathcal{I}_7
\cr && +123480 \mathcal{I}_8)\,,
\cr &&
\cr && S_{-1}=\frac{25}{8} \pi ^2 (12 \mathcal{I}_0+6999 \mathcal{I}_1+379278 \mathcal{I}_2+972310 \mathcal{I}_3-184455212 \mathcal{I}_4-3753014472 \mathcal{I}_5
\cr && -36152126988 \mathcal{I}_6-212468552382 \mathcal{I}_7-837434528352 \mathcal{I}_8-2318322569528 \mathcal{I}_9
\cr && -4609877097728 \mathcal{I}_{10}-6626423377980 \mathcal{I}_{11}-6831406410864 \mathcal{I}_{12}-4926403116096 \mathcal{I}_{13}
\cr && -2359679148288 \mathcal{I}_{14}-674380586880 \mathcal{I}_{15}-87016849920 \mathcal{I}_{16})-\frac{75}{8} \pi ^2 \log2\,(7 \mathcal{I}_0+2540 \mathcal{I}_1
\cr && +82440 \mathcal{I}_2-2164092 \mathcal{I}_3-117428188 \mathcal{I}_4-1968851772 \mathcal{I}_5-17936516924 \mathcal{I}_6-103383558648 \mathcal{I}_7
\cr && -404896847592 \mathcal{I}_8-1119462183840 \mathcal{I}_9-2227268584128 \mathcal{I}_{10}-3204982912320 \mathcal{I}_{11}
\cr && -3307460647680 \mathcal{I}_{12}-2386946580480 \mathcal{I}_{13}-1143848724480 \mathcal{I}_{14}-326972405760 \mathcal{I}_{15}
\cr && -42189987840 \mathcal{I}_{16})\,.
\eea
It is not hard to verify the Residue relation from $S_{-2}$ above.
\end{itemize}

\subsubsection{$\Delta_L=4$}
In the case of $\Delta_L=4$, the mixing happens to sub-sub-leading-twist.
\begin{itemize}
\item[1.] Leading-twist
\bea
&& S_{-1}=\frac{1}{2} \pi ^2 (3 \mathcal{I}_0+2034 \mathcal{I}_1+184115 \mathcal{I}_2+5546904 \mathcal{I}_3+79479952 \mathcal{I}_4+650656672 \mathcal{I}_5
\cr && +3374106104 \mathcal{I}_6+11765981920 \mathcal{I}_7+28516757080 \mathcal{I}_8+48712874000 \mathcal{I}_9+58497502280 \mathcal{I}_{10}
\cr && +48347712000 \mathcal{I}_{11}+26186462400 \mathcal{I}_{12}+8367004800 \mathcal{I}_{13}+1195286400 \mathcal{I}_{14})-3 \pi ^2 \log2\,(\mathcal{I}_0
\cr && +628 \mathcal{I}_1+50980 \mathcal{I}_2+1429728 \mathcal{I}_3+19711944 \mathcal{I}_4+158456832 \mathcal{I}_5+815776944 \mathcal{I}_6+2839435680 \mathcal{I}_7
\cr && +6884892120 \mathcal{I}_8+11774800800 \mathcal{I}_9+14156731680 \mathcal{I}_{10}+11711078400 \mathcal{I}_{11}+6346636800 \mathcal{I}_{12}
\cr && +2028364800 \mathcal{I}_{13}+289766400 \mathcal{I}_{14})\,.
\eea

\item[2.] Sub-leading twist
\bea
&& S_{-1}=\frac{1}{2} \pi ^2 (9 \mathcal{I}_0+3960 \mathcal{I}_1+179922 \mathcal{I}_2+814558 \mathcal{I}_3-49410876 \mathcal{I}_4-911304904 \mathcal{I}_5
\cr && -7596112752 \mathcal{I}_6-38105022160 \mathcal{I}_7-126821192680 \mathcal{I}_8-292642807520 \mathcal{I}_9-476503711600 \mathcal{I}_{10}
\cr && -546738918320 \mathcal{I}_{11}-432930876000 \mathcal{I}_{12}-225268209600 \mathcal{I}_{13}-69326611200 \mathcal{I}_{14}-9562291200 \mathcal{I}_{15})
\cr && -6 \pi ^2 \log2\,(\mathcal{I}_0+368 \mathcal{I}_1+13000 \mathcal{I}_2-84872 \mathcal{I}_3-7570136 \mathcal{I}_4-117039648 \mathcal{I}_5-932926096 \mathcal{I}_6
\cr && -4615513760 \mathcal{I}_7-15307215000 \mathcal{I}_8-35325058560 \mathcal{I}_9-57585511200 \mathcal{I}_{10}-66154085760 \mathcal{I}_{11}
\cr && -52432665600 \mathcal{I}_{12}-27298252800 \mathcal{I}_{13}-8403225600 \mathcal{I}_{14}-1159065600 \mathcal{I}_{15})\,.
\eea

\item[3.] Mixed
\bea
&& S_{-2}=\frac{3}{4} \pi ^2 (\mathcal{I}_0+440 \mathcal{I}_1+16280 \mathcal{I}_2+171680 \mathcal{I}_3+778840 \mathcal{I}_4+1792000 \mathcal{I}_5+2198000 \mathcal{I}_6
\cr && +1372000 \mathcal{I}_7+343000 \mathcal{I}_8)\,,
\cr &&
\cr && S_{-1}=\frac{1}{8} \pi ^2 (63 \mathcal{I}_0+19296 \mathcal{I}_1+647940 \mathcal{I}_2+9031736 \mathcal{I}_3+199507148 \mathcal{I}_4+3890478864 \mathcal{I}_5
\cr &&+42847963888 \mathcal{I}_6+285325317280 \mathcal{I}_7+1245992534280 \mathcal{I}_8+3751666766400 \mathcal{I}_9
\cr && +8001253401120 \mathcal{I}_{10}+12205757188800 \mathcal{I}_{11}+13244672767200 \mathcal{I}_{12}+9987643344000 \mathcal{I}_{13}
\cr && +4975891200000 \mathcal{I}_{14}+1472592844800 \mathcal{I}_{15}+196026969600 \mathcal{I}_{16})-\frac{3}{4} \pi ^2 \log2\,(13 \mathcal{I}_0+4040 \mathcal{I}_1
\cr && +181496 \mathcal{I}_2+3887328 \mathcal{I}_3+73926184 \mathcal{I}_4+1114649088 \mathcal{I}_5+10983854896 \mathcal{I}_6+70165706720 \mathcal{I}_7
\cr && +302034903480 \mathcal{I}_8+905878243200 \mathcal{I}_9+1931862606720 \mathcal{I}_{10}+2950250822400 \mathcal{I}_{11}
\cr && +3205199088000 \mathcal{I}_{12}+2419240857600 \mathcal{I}_{13}+1205970124800 \mathcal{I}_{14}+356992204800 \mathcal{I}_{15}
\cr && +47521689600 \mathcal{I}_{16})\,,
\eea
where $S_{-2}$ can be used to verify the Residue relation easily.
\end{itemize}

\end{document}